\documentclass[aip,reprint]{revtex4-2}

\usepackage{hyperref}
\hypersetup{colorlinks=true,linkcolor=black,citecolor=black,urlcolor=blue,hyperfootnotes=true}
\usepackage{graphicx}
\usepackage{amsmath}

\newcommand{\bose}{\bar{n}}
\draft 

\begin{document}

\title[Quantum thermodynamics near the border of a Bose gas]%
{Quantum thermodynamics near the border of a one-dimensional Bose gas}

\author{Carsten Henkel}
\email[]{henkel@uni-potsdam.de}
\affiliation{University of Potsdam, Institute of Physics and Astronomy, 
Karl-Liebknecht-Str. 24/25, 14476 Potsdam, Germany}


\begin{abstract}
We consider an ultracold Bose gas in a half-open potential well, otherwise confined to a quasi-one-dimensional geometry. 
The Bogoliubov equations for its elementary excitations are solved in the continuous spectrum to explore the particle and energy content near the edge of the gas, beyond the often applied local density approximation.
In particular, gradients in the condensate density enhance density-dominated excitations in the border region. 
We discuss excess (missing) particles and their energy by comparing to suitable reference solutions for quasi-homogeneous systems.
The density profile near the edge shows no Friedel oscillations, but a dipolar feature from the spill-out of thermally excited particles.
The calculations are performed in the grand-canonical ensemble and in the thermodynamic limit.
\end{abstract}


\maketitle 

\section{Introduction}

Ultracold atomic gases provide well-controlled model systems to study many-body physics where temperature, interaction strength, and dimensionality/confinement can be varied over a wide range.\cite{PitaevskiiStringari, ProukakisBook}
A wealth of information is available from spatially resolved probes of the particle density, leading to the equation of state and spatial correlations, to name a few typical observables.\cite{Giorgini00, Kheruntsyan05, Henkel17b}
In this paper, we extend previous work on a quasi-one-dimensional Bose gas and its spatial cross-over from high to low densities.\cite{AlKhawaja02b, Diallo15a, Henkel25b, Henkel25c} 
The focus is on the edge of an anisotropically trapped system: in two dimensions a strong confinement is applied, while in the third, axial dimension the system is closed by a potential barrier with a turning point.
We consider different barrier types that range from a soft (penetrable) potential to a hard wall, where a Dirichlet boundary condition is applied to the condensate order parameter and to its elementary excitations.
Our calculations address the degenerate and ideal Bose gas regimes that can be handled with temperature-dependent mean-field theory \cite{Mora03, BouchouleChapter11, Kerr24},
and provide numerically exact solutions across the turning point.\cite{Diallo15a,Henkel25c}
In this way, we avoid relying on the local density approximation.

\begin{figure}[tbh]
\centerline{%
\hspace*{05mm}
\includegraphics*[height=0.4\columnwidth]{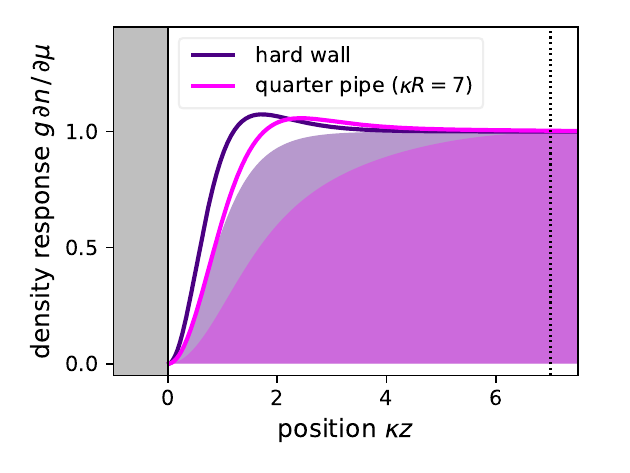}
\hspace*{-3.5mm}%
\includegraphics*[height=0.4\columnwidth]{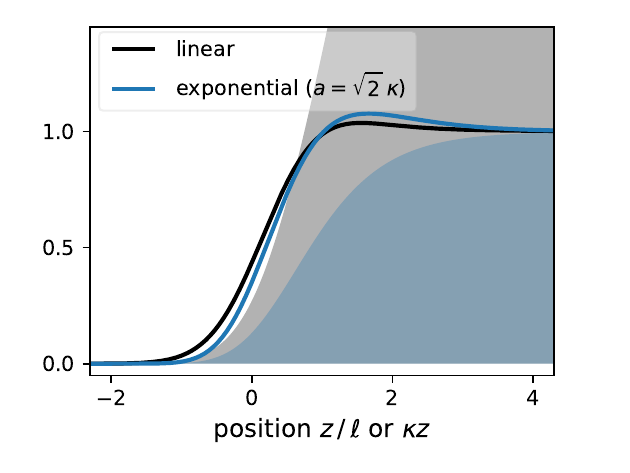}
}

\caption[]{%
Kaleidoscope of condensate edges considered in this paper. 
Left panel: hard-wall potentials, right panel: soft potentials.
The coloured shaded areas illustrate the density $n(z) = |\phi(z)|^2$ of the (quasi-)condensate mean field.
The solid lines provide the response $\partial n / \partial \mu$ to a change in the chemical potential, related to the (local) compressibility.
Parameters: hard-wall potential $V = 0$, ``quarter-pipe'' potential connecting smoothly a hard wall at $z = 0$ to a flat bottom at $z \ge R$ (dotted line), exponential $V(z) = \mu\exp(-a z)$ and linear $V(z) = \mu - F z$ potential. 
See Table\:\ref{t:units} for the potential parameters and the length scales $1/\kappa$ (healing length) and $\ell$ (linear potential).
The turning point is at $z = 0$ in all cases, although the boundary conditions there depend on the type of potential (Dirichlet $\psi(0) = 0$ for the hard-wall potentials in the left panel).
}
\label{fig:sketches}
\end{figure}

The companion paper of Ref.\:\onlinecite{Henkel25c} has outlined a scheme to solve the Gross-Pitaevskii and Bogoliubov equations for the condensate and its elementary excitations.\cite{PitaevskiiStringari}
The scheme avoids instabilities that arise in the continuous spectrum because the characteristic equation for the Bogoliubov problem is of fourth order and contains exponentially growing solutions.
A key feature of this scheme is the separation of the solutions into phase and density modes which are coupled by the density gradient of the condensate, particularly strong in the boundary region.
A qualitative sketch is provided in Fig.\:\ref{fig:sketches} where the condensate density and its response to a change in the chemical potential are shown for four different model potentials.
See Table~\ref{t:units} for details and parameters.

In this paper, the spatial profiles of particle and energy densities are computed both at zero and finite temperature.
In the former limit, studied recently in Ref.\:\onlinecite{Henkel25b}, only quantum fluctuations of the second-quantised particle wave field play a role and ``deplete'' the condensate.
This may be understood from (virtual) scattering events that promote a pair of particles to elementary excitations outside the condensate.\cite{PitaevskiiStringari}
These interactions are diagonalised with the help of the celebrated Bogoliubov transformation whose eigenstates describe the elementary excitations of the condensate. 
At low energies, they are mostly collective and describe fluctuations in the condensate phase.
This scheme acquires a richer phenomenology in the boundary region where excitations of the phase and density type are strongly coupled.\cite{Diallo15a, Henkel25c} 
We observe in particular a ``dipolar'' feature in the finite-temperature density profile where more particles are found beyond the turning point (forming a dilute Bose gas), while in the dense phase, the total particle density is reduced. 
The latter is due to the back action of the non-condensed particles on the condensate wave function. 
This can be qualitatively captured by a Hartree-Fock-Bogoliubov equation that reproduces the Bogoliubov theory extended to quasi-condensates of \citet{Mora03}, allowing for strong phase, but weak density fluctuations. 

Quite generally, the approach followed here is similar to calculations of the electronic density at the surface of a metal.\cite{Liebsch_Book, Toscano15} 
Indeed, a more or less smooth potential provides the confinement of the mobile charges to the ionic cores.
This is sometimes modelled from a step-like positive charge density, in the so-called jellium approximation, corrected consistently for the effect of exchange.\cite{Heitler_1927b, Dirac_1930, Lang69, Kenner72, Liebsch_Book, Yan15b}
The electron density profile is built from one-particle wave functions in this potential. 
Although the parameter ranges of bosonic and fermionic systems are vastly different (the latter may reach the size of a neutron star), they have one qualitative feature in common, namely Friedel oscillations of the particle density in the vicinity of the surface.\cite{Lang69}
For the Bose gas, the latter simplify to the dipolar feature mentioned above, however, because there is nothing like a sharp Fermi edge.
The energy localised in the border region may be interpreted as a surface tension.\cite{Schaeybroeck08} 
It can be defined unambiguously by comparing to a homogeneous system of similar size whose termination is shifted relative to the nominal turning point.
Similarly, a displaced reference plane has been put forward when defining particle-surface distances in van der Waals interactions.\cite{Zaremba76}

In Sec.\:\ref{s:Bogoliubov-framework}, we review the framework of extended Bogoliubov theory that we use.\cite{Mora03}
The energy density of the inhomogeneous Bose gas is discussed in Sec.\:\ref{s:energy-density} and evaluated for its ground state, including condensate depletion. 
In Sec.\:\ref{s:depletion-thermal}, the particle density is worked out. 
We explain how the Bogoliubov theory for quasi-condensates avoids the infrared singularities that often plague mean field approaches at finite temperature due to the gapless excitation spectrum.\cite{AlKhawaja02b, Henkel17b}
The appendices provide rather technical material, in particular corrections to the condensate density that appear in the grand-canonical ensemble. These can be worked out in some detail, but actually vanish in the strict thermodynamic limit of a half-open system with a continuous spectrum.

\begin{table*}[t!bh]
\begin{center}
\begin{tabular}{l|llll}
& hard wall & quarter pipe & exponential & linear
\\
\hline
potential $V(z)$ &
$0$ & $\mu - \mu \sqrt{(2 - \frac zR) \frac zR}$ & 
$\mu\,\exp(- a z)$ & $\mu - F z$
\\
length scale(s) &
$\frac{1}{\kappa} = (\frac{\hbar^2}{\mu M})^{1/2}$ &
range $R$ &
$1/a$ &
$\ell = (\frac{\hbar^2}{2 M F})^{1/3}$
\\
value & $230\,{\rm nm}$ 
& $R = 1 \ldots 1.8\,\mbox{µ}{\rm m}$
& $50\ldots 200\,{\rm nm}$ 
& $300\,{\rm nm}$
\\
parameters & $\mu = 100\,{\rm nK}$
& 
& 
& $F/M = 9.81\,{\rm m/s}^2$
\\
numerics
& $\mu = 1$
& $\kappa R = 2 \ldots 10$
& $\kappa/a = 0.24 \ldots 2.4$
& $\mu = 0$, $F = 1 = \ell$
\end{tabular}
\end{center}
\caption[]{%
Characteristic units and parameters. 
The numbers (atom mass $M$, resonant wavelength) are evaluated for the Rb-87 atom.
The quarter-pipe potential continues as $V(z) = 0$ for $z \ge R$.
The exponential potential can be implemented with a blue-detuned light field created by total internal reflection (evanescent wave). 
Its length scale $1/a$ is of the order of the Rb D2 wavelength $780\,{\rm nm}$, but depends on optical parameters.
The energy scale in the linear potential is $F \ell \approx 31\,{\rm nK}$ for standard gravity on Earth.
For the numerics, we adopt the typical value $g \approx 2\,{\rm nK}\,\mbox{µ}{\rm m}$ for the interaction parameter. 
This value depends on details of the confinement in the directions transverse to the $z$-axis and can be tuned over a fairly wide range using Feshbach resonances in a magnetic field because it is proportional to the s-wave scattering length.\cite{Olshanii98, BouchouleChapter11, DrakeHandbook}
The numerics uses units with $\hbar^2/2M = 1$.
}
\label{t:units}
\end{table*}

\section{Extended mean-field theory framework}
\label{s:Bogoliubov-framework}

In the Bogoliubov theory extended to inhomogeneous quasi-condensates, the total field energy is given by the spatial integral of the grand-canonical operator \cite{Mora03}
\begin{equation}
K = - \frac12 
\Big(
\Psi^\dag \frac{ {\rm d}^2\Psi }{ {\rm d}z^2 } 
+
\frac{ {\rm d}^2\Psi^\dag }{ {\rm d}z^2 } \Psi
\Big)
+ (V - \mu) \, \Psi^\dag \Psi + \frac{g}{2} \Psi^{\dag2} \Psi^2
\label{eq:gcan-Hamiltonian}
\end{equation}
where units with $\hbar^2 / 2M = 1$ are used, and $g > 0$ parametrises the interaction energy when two particles are located at the same position (contact potential\cite{PitaevskiiStringari}).
It is somewhat conventional how to write the first term (kinetic energy density).
With suitable boundary conditions, the conventions are equivalent once the spatial integration is performed. 
We adopt this form because it is locally Hermitean.
See Table~\ref{t:units} for typical potentials and physical parameters.
Due to their low density (few tens of particles per micron), temperatures in ultracold, degenerate Bose gases are in the $100\,{\rm nK}$ range.

\subsection{Order zero: condensate}

The basic assumption of the \citet{Mora03} scheme is to expand the energy density $\langle K \rangle = e(z) = e^{(0)} + e^{(2)} + \ldots$ up to the second order in fluctuations around the condensate and minimise.
The condensate is found in the zeroth order by approximating the operator $\Psi$ by a complex field $\phi$ in Eq.\,(\ref{eq:gcan-Hamiltonian}).
We thus get the stationary Gross-Pitaevskii equation
\begin{equation}
- \frac{ {\rm d}^2\phi }{ {\rm d}z^2 } 
+ (V - \mu) \, \phi + g \phi^3 = 0
\label{eq:GPe-1}
\end{equation}
Without loss of generality, we have assumed here that $\phi$ is real.
The physically acceptable solution behaves according to the Thomas-Fermi approximation on the dense side of the turning point
\begin{equation}
z \to +\infty: \quad
g \phi^2(z) \approx \mu - V(z)
\label{eq:Thomas-Fermi-density-1}
\end{equation}
where the second derivative is neglected (slowly varying confinement).
For a condensate confined by a hard wall, Eq.\,(\ref{eq:GPe-1}) is exactly solved by\cite{PitaevskiiStringari}
\begin{equation}
g \phi^2(z) = \mu\, \tanh \kappa z
\label{eq:hard-wall-condensate}
\end{equation}
with the ``healing'' length $1/\kappa \sim \mu^{-1/2}$ of Table~\ref{t:units}.

Using Eq.\,(\ref{eq:GPe-1}) in $\langle\hat{K}\rangle$, we get for the energy density in zeroth order
\begin{equation}
e^{(0)}(z) = - 
\phi \,\frac{ {\rm d}^2\phi }{ {\rm d}z^2 }
+  (V - \mu) \, \phi^2 + \frac{g}{2} \phi^4 
= - \frac{g}{2} \phi^4 
\label{eq:energy-density-0}
\end{equation}
The behaviour of these quantities in the border region has been discussed in Ref.\:\onlinecite{Henkel25b}.
A brief summary and the extension to the confinement in a linear potential (last column of Tab.\:\ref{t:units}) is given in Sec.\:\ref{s:gd-state-energy} below.

\subsection{Order two: fluctuations}

The first-order correction to $\langle K \rangle$ vanishes by the choice of $\phi$ minimising the energy.
In the second order, the expansion of the energy density~(\ref{eq:gcan-Hamiltonian}) contains two c-number contributions\cite{Henkel25b}\textsuperscript{,}
\footnote{%
Equation~(\ref{eq:gd-state-energy-density}) differs slightly from Eq.\,(113) by \citet{Mora03} because we do not follow their steps involving the spatial projection orthogonal to the condensate mode. 
}
\begin{equation}
e^{(2)}(z) = 
- \frac{\mu'_0}{2} |\phi(z)|^2
- \int\!\frac{{\rm d}E}{\pi} E |v(z; E)|^2
\label{eq:gd-state-energy-density}
\end{equation}
The first term arises from the grand-canonical fluctuation in the number of (condensate) particles and scales with the derivative 
\begin{equation}
\mu'_0 =
  \left(\frac{ \partial N_0 }{ \partial \mu }\right)^{-1}
\label{eq:def-mu-prime-0}
\end{equation}
One considers here the system confined to a large, but finite size $L$ so that the spatial integral of the condensate density $\phi^2(z)$ converges to $N_0$. 
The second term in Eq.\,(\ref{eq:gd-state-energy-density}) arises from quantum fluctuations of the particle density: collisions between condensate particles can virtually promote particles to modes with a higher energy.\cite{AlKhawaja02b, Mora03}
This effect is also known as depletion of the condensate.\cite{PitaevskiiStringari}

The field fluctuations are described by the operator $\hat{\psi} = \Psi - \phi$ which is expanded as
\begin{align}
\hat{\psi}(z) &= 
{-} {\rm i} \hat{Q} \, \phi(z) + \hat{P} \, \phi_a(z) 
+ \hat{B}(z)
\nonumber\\
\hat{B}(z) &= 
\int_0^\infty\!\frac{ {\rm d}E }{ \mbox{\small${\sqrt{\pi}}$} } 
\left[ u(z; E)\, \hat{b}(E) + v(z; E)\, \hat{b}^\dag(E)\right]
\label{eq:fluctuation-operator}
\end{align}
In the first two terms of $\hat{\psi}$, the Hermitean operators $\hat{Q}$, $\hat{P}$ describe phase and particle number fluctuations in the condensate, respectively, while $\hat{B}$ contains the standard Bogoliubov quasi-particle excitations, see Eq.\,(\ref{eq:bbdag-commutator}) below.
In order to secure the standard commutator $[\hat{Q}, \hat{P}] = {\rm i}$, the adjoint mode function $\phi_a$ is defined by the variation of the condensate field with respect its particle number\cite{Castin98, Dziarmaga04}
\begin{equation}
\phi_a 
= \frac{ \partial \phi }{ \partial N_0 }
= \mu'_0 
  \frac{ \partial \phi }{ \partial \mu }
\label{eq:def-phi-a}
\end{equation}
It is convenient to extract the factor $\mu'_0$ because the derivative $\partial \phi / \partial \mu$ has a well-defined thermodynamic limit (e.g., $N_0$ and $L \to \infty$ at fixed $n_0 = N_0/L$ for a potential with a flat bottom).
The adjoint mode is computed by taking the $\mu$-derivative of the Gross-Pitaevskii equation~(\ref{eq:GPe-1}) which gives
\begin{align}
H_3 \frac{ \partial \phi }{ \partial \mu }
& = \phi
\nonumber\\
H_3 &=
- \frac{ {\rm d}^2 }{ {\rm d}z^2 } + 
V(z) - \mu + 3 g \phi^2(z)
\label{eq:H3-chi}
\end{align}
Since the potential in $H_3$ scales as $2 g \phi^2$ in the Thomas-Fermi limit~(\ref{eq:Thomas-Fermi-density-1}) (neglecting the derivatives), a simple approximate solution of Eq.\,(\ref{eq:H3-chi}) is
\begin{equation}
\frac{\partial\phi}{\partial\mu} 
\approx
\frac{ 1 }{ 2 g \, \phi }
\label{eq:solution-phi-a}
\end{equation}
valid in the bulk of the system for real and positive $\phi$.
The derivative $\partial |\phi|^2/\partial\mu$ thus saturates to the constant value $1 / g$. 
This even holds for the linear potential, as illustrated in Fig.\:\ref{fig:sketches}.
For a more accurate calculation, we use a finite-difference scheme and observe that the lowest eigenvalue of the Hamiltonian $H_3$ is strictly positive.
A numerical check is provided by the identity
\begin{equation}
\int^{L}\!{\rm d}z \, \phi \, \phi_a = \frac{1}{2}
\label{eq:}
\end{equation}
that follows from differentiating the integral that defines the condensate particle number, with respect to $N_0$.

The continuum of elementary excitations in Eq.\,(\ref{eq:fluctuation-operator}) is characterised by a pair $u, v$ of mode functions and bosonic operators $\hat{b}(E)$, $\hat{b}^\dag(E)$ that annihilate and create non-condensate quasiparticles with energy $E > 0$ measured relative to the chemical potential $\mu$.
Their properties, in particular for a continuous spectrum, have been discussed in Refs.\:\onlinecite{Diallo15a, Henkel25c}, and the main features are summarised here.
The quasiparticle operators obey the commutation relations
\begin{equation}
\left[ \hat{b}(E) \,,\, \hat{b}^\dag(E') \right] = \delta(E - E')
\label{eq:bbdag-commutator}
\end{equation}
and commute with the operators $\hat{Q}$, $\hat{P}$ mentioned above.
The ``particle'' and ``hole'' wave functions $u$ and $v$ are the solutions of the Bogoliubov equations\cite{PitaevskiiStringari}
\begin{align}
\phantom{-} E \, u &=
H_2 u 
+ g \phi^2(z) \, v 
\nonumber\\
- E \, v &=
H_2 v + g \phi^2(z) \, u 
\label{eq:BdG-1}
\\
H_2 
&= - \frac{ {\rm d}^2 }{ {\rm d}z^2 } + V - \mu  + 2 g \phi^2(z)
\label{eq:def-H2}
\end{align}
The energy integration measure ${\rm d}E / \sqrt{\pi}$ in Eq.\,(\ref{eq:fluctuation-operator}) and the commutator in Eq.\,(\ref{eq:bbdag-commutator}) require for consistency the following normalization of the continuum mode functions $u(z; E)$, $v(z; E)$
\begin{equation}
\int\!\frac{{\rm d}z}{\pi} 
\left[
u(z; E) \, u(z; E') - v(z; E) \, v(z; E')
\right]
= \delta( E - E')
\label{eq:u-v-orthogonality}
\end{equation}
A similar orthogonality holds with respect to the condensate mode:
\begin{equation}
\int\!{\rm d}z
\left[
u(z; E) + v(z; E) 
\right]
\phi(z) 
= 0
\label{eq:}
\end{equation}
This is in fact a special case of Eq.\,(\ref{eq:u-v-orthogonality}), because the pair $u, v = \phi, -\phi$ solves Eqs.\,(\ref{eq:BdG-1}) with $E = 0$.

The normalization~(\ref{eq:u-v-orthogonality}) is secured by a proper amplitude for the mode functions in the asymptotic region $z \to \infty$, as discussed by \citet{Diallo15a}.
Consider first the case that the potential tends to a constant (set to zero). 
The Bogoliubov modes asymptotically become plane waves with a wavenumber $k$ set by the dispersion relation
\begin{equation}
E = |k| \sqrt{ 2 \mu + k^2 }
\label{eq:Bogoliubov-dispersion-II}
\end{equation}
They are normalised according to
\begin{eqnarray}
& u(z; E) & \to \phantom{-} 
\cos(\theta/2)
\left[ j_E(z) \cos\delta - y_E(z) \sin\delta \right]
\nonumber\\
& v(z; E) & \to 
-\sin(\theta/2)
\left[ j_E(z) \cos\delta - y_E(z) \sin\delta \right]
\,,
\label{eq:uv-normalisation-flat-bottom}
\end{eqnarray}
where the angle $\theta$ is defined by
\begin{equation}
\tan\theta = \frac{ \mu }{ E }
\label{eq:theta-angle-hw}
\end{equation}
The (energy-dependent) phase shift $\delta$ characterises the node positions of the standing waves, measured relative to the pair of reference solutions
\begin{equation}
j_E(z) = \frac{ \sin kz }{ \sqrt{k} }
\,,
\quad
y_E(z) = -\frac{ \cos kz }{ \sqrt{k} }
\label{eq:def-jy-flat-bottom}
\end{equation}
The function $j_E(z)$ satisfies a Dirichlet boundary condition at $z = 0$ as if the potential barrier were replaced by a hard wall there. 
In the particular case of a hard-wall confinement ($V = 0$ for $z > 0)$, $\tan\delta = 2 \kappa / k$ exactly\cite{Henkel25c}, with $\kappa = \sqrt{\mu/2}$ the inverse healing length introduced in Table\:\ref{t:units}.

The normalization~(\ref{eq:uv-normalisation-flat-bottom}) can be generalised to a half-open potential that is not constant. 
In the particular case of a linear potential $V = \mu - Fz$, the phase $\theta = \theta(z)$ becomes position dependent according to
\begin{equation}
\tan\theta(z) = F z / E
\label{eq:def-theta-linear}
\end{equation}
The reference solutions in Eq.\,(\ref{eq:def-jy-flat-bottom}) are replaced by \cite{Diallo15a, Henkel25c}
\begin{align}
j_E(z) & = \sqrt{ \pi z } \, J_0( E \sqrt{ 2 z } / F \ell^{3/2} )
\nonumber\\
y_E(z) & = \sqrt{ \pi z } \, Y_0( E \sqrt{ 2 z } / F \ell^{3/2} )
\label{eq:Bessel-asymptote-linear}
\end{align}
with the ordinary zeroth-order Bessel functions $J_0$, $Y_0$.
The characteristic length $\ell$ is linked to the force in the linear potential, see last column in Tab.\:\ref{t:units}.
--
For both types of mode continua, the normalization required by Eq.\,(\ref{eq:u-v-orthogonality}) is ensured because of a unit Wronskian between reference solutions:
\begin{equation}
W[j_E, y_E] = j_E \frac{ d y_E }{ d z } - y_E \frac{ d j_E }{ d z } = 1
\,,
\label{eq:Wronskian-Bessel}
\end{equation}
For the Bessel functions, this follows immediately from the identity $J_0(x) Y'_0(x) - J'_0(x) Y_0(x) = 2/(\pi x)$ [Eq.\,(10.5.2) in Ref.\:\onlinecite{DLMF}].
The general proof is reviewed in Appendix~A of Ref.\:\onlinecite{Diallo15a}.
For a discussion of the phase shift $\delta$, see Ref.\:\onlinecite{Henkel25c}.

\subsection{Energy}

Finally, the \emph{energy} of the fluctuations that defines the thermodynamic equilibrium of the quasi-condensate (within second-order accuracy) is given by the familiar quadratic form
\begin{equation}
H^{(2)} = \frac{\mu'_0}{2} \hat{P}^2 + 
\int_0\!{\rm d}E \, E \, \hat{b}^\dag(E) \, \hat{b}(E)
\label{eq:H2}
\end{equation}
\citet{Mora03} construct the thermal equilibrium for fluctuations based on this Hamiltonian.
This yields vanishing mean values for $\hat{P}$ and $b(E)$, and 
\begin{eqnarray}
\phantom{\langle \hat{b}^\dag(E) \, \hat{b}(E') \rangle }%
\llap{%
$\langle \hat{P}^2 \rangle$
} 
& = & T / \mu'_0
\,,
\label{eq:P-square-average}
\\
\langle \hat{b}^\dag(E) \, \hat{b}(E') \rangle 
& = &
\bose(E) \, \delta( E - E' )
\label{eq:b-bdag-average}
\\
\langle \hat{b}(E) \, \hat{b}^{\dag}(E') \rangle 
& = &
\left[ 1 + \bose(E) \right]
\delta( E - E' )
\nonumber
\end{eqnarray}
where the Bose function is $\bose(E) = 1/[\exp(E / T)- 1]$ with $k_B = 1$.

The local version of Eq.\,(\ref{eq:H2}) is given by the energy density operator [Eqs.\,(10, A.6) of Ref.\:\onlinecite{Henkel25b}]:
\begin{eqnarray}
\hat{K}^{(2)} & = & \mu'_0 \hat{P}^2 \phi_a \phi 
\label{eq:expand-H2-step-1}
\\
&& +
\int_0\!\frac{{\rm d}E\, {\rm d}E'}{\pi} 
\frac{E}{2} 
\left[:\!
\big( u^{\prime*} \, \hat{b}^{\prime\dag} + v^{\prime*} \, \hat{b}' \big)
\big( u \, \hat{b} - v \, \hat{b}^\dag \big)
\right.
\nonumber\\
&& \hphantom{ +
\int_0\!\frac{{\rm d}E\, {\rm d}E'}{\pi} \frac{E}{2}
}
\left. {} + 
\big( u^* \, \hat{b}^\dag - v^* \, \hat{b} \big)
\big( u' \, \hat{b}' + v' \, \hat{b}^{\prime\dag} \big)
\!:\right]
\nonumber
\end{eqnarray}
where $\phi_a$ is the adjoint mode introduced in Eq.\,(\ref{eq:def-phi-a}).
The energy parameter $E$ has been suppressed from $u, v, b, b^\dag$ for simplicity; primed quantities correspond to the energy $E'$. 
The colons ${:} \cdots {:}$ denote normal ordering of the $\hat{b}^\dag, \hat{b}$ operators.
The second term of the quantum (ground state) energy in Eq.\,(\ref{eq:gd-state-energy-density}) can be recognised in Eq.\,(\ref{eq:expand-H2-step-1}) when the operator products $-v^{\prime *} v \, \hat{b}' \hat{b}^\dag$ and $-v^{*} v' \, \hat{b} \, \hat{b}^{\prime \dag}$ are put into normal order.
(The first term arises from ordering the condensate operators $\hat{Q}, \hat{P}$.)

\begin{figure}[btph]
   \centering
   \includegraphics[width=0.85\columnwidth]{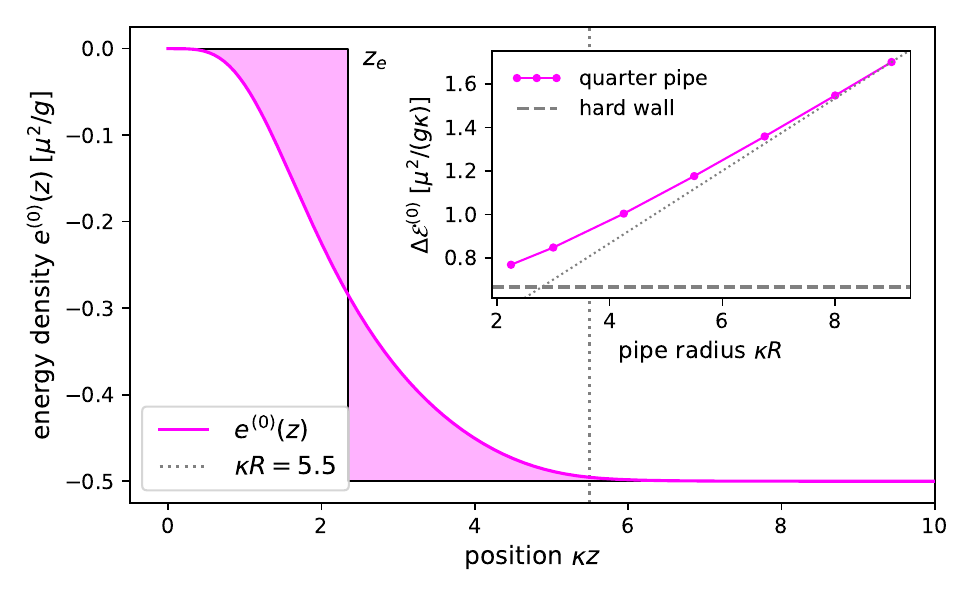} 
   \caption[]{%
   Ground state energy density $e^{(0)}(z)$ due to the condensate in the quarter-pipe potential. 
   The position $z_e$ marks the abrupt termination of a homogeneous system with the same bulk density and (total) ground state energy (the shaded areas add to zero).
   Its product with the energy density $-\mu^2/(2g)$ gives the condensate contribution $\Delta {\cal E}^{(0)}$ to the surface energy. 
   The inset illustrates the scaling with the “radius” $R$ of the quarter-pipe potential, the dotted line is the prediction of the Thomas-Fermi approximation.}
   \label{fig:surface-energy-qpipe}
\end{figure}

\begin{table*}[tbhp]
\begin{center}
   \begin{tabular}{@{} l|l l@{$\,,\ $}l @{\quad} l@{$\,,\ $}l @{\quad} l} 
    \hline
	Potential
	& hard wall
	& \multicolumn{2}{c}{quarter-pipe}
	& \multicolumn{2}{c}{exponential}
	& linear
	\\
    \hline
	$z_e$ (energy)
	& $4/(3\kappa)$ 
	& $1.70 / \kappa$	& $\kappa R = 3$
	& $1.46 / \kappa$	& $a = 1.41\,\kappa$
	& $1.28\,\ell$
	\\
	&
	& $2.35 / \kappa$	& $\kappa R = 5.5$
    & $1.82 / \kappa$	& $a = 1 \, \kappa$ 
    \\
	&
    & $3.09 / \kappa$ 	& $\kappa R = 8$ 
    & $2.88 / \kappa$	& $a = 0.57\,\kappa$
	\\	
	\hline
	$z_d$ (density)
	& $1/\kappa$ 
	& $1.29 / \kappa$	& $\kappa R = 3$
	& $0.95 / \kappa$	& $a = 1.41\,\kappa$
	& $0$
	\\
	&
	& $1.74 / \kappa$	& $\kappa R = 5.5$
	& $1.19 / \kappa$	& $a = 1 \, \kappa$
    \\
	&
    & $2.24 / \kappa$ 	& $\kappa R = 8$ 
    & $2.46 / \kappa$	& $a = 0.42\,\kappa$
	\\	
    \hline
   \end{tabular}
\end{center}
\caption[]{%
Depletion lengths $z_e$, $z_d$ for ground-state surface energy $\Delta {\cal E}^{(0)}$ and condensate particle number.
According to Fig.\:\ref{fig:surface-energy-qpipe}, the integral of the ground state energy scales according to 
${\cal E}^{(0)}(L) 
= - (L - z_e) \mu^2/(2g) 
= - L \mu^2/(2g) + \Delta {\cal E}^{(0)}$ 
[Eq.\,(\ref{eq:scaling-gd-state-energy-flat})].
For the particle number, the scaling is
$N^{(0)}(L) = (L - z_d) \mu/g$, 
see Appendix~\ref{a:boundary-corrections}.
Exact results for the hard-wall potential.
For the linear potential, the scaling of ${\cal E}^{(0)}(L)$ is given in Eq.\,(\ref{eq:scaling-gd-state-energy-linear}); $N^{(0)}(L)$ is discussed in Appendix~\ref{a:dipole}.
}
\label{t:ze-0-surface-energy}
\end{table*}

\section{Equilibrium energy density}
\label{s:energy-density}

\subsection{Ground state energy}
\label{s:gd-state-energy}

\subsubsection{Condensate contribution}

The spatial profile of the condensate ground state energy is illustrated in Fig.\:\ref{fig:surface-energy-qpipe} for a potential barrier with a flat bottom.
The energy density saturates to $-\mu^2 / 2g$ in the bulk.
Subtracting this limit and integrating, one may define the boundary energy, similar to a surface tension, by the integral\cite{Schaeybroeck08}
\begin{equation}
\Delta {\cal E}^{(0)} = 
- \frac{1}{2} \int\limits^{\infty}_{}\!{\rm d}z\, 
\Big(
g\phi^4(z) - \frac{\mu^2}{g} \Theta(z) \Big)
\label{eq:split-total-energy-0}
\end{equation}
where $\Theta(z)$ is the step function.
As can be seen in Fig.\:\ref{fig:surface-energy-qpipe}, the surface energy is proportional to the size of the transition region near the turning point, times the bulk energy density (inset, dotted line).
A natural parametrization is thus to express the energy ${\cal E}^{(0)}(L)$ up to the finite, but large position $z = L \gg 1/\kappa$ as an effective length $z_e$ (subscript $e$ for ``energy'')
\begin{equation}
{\cal E}^{(0)}(L) = - \frac{ \mu^2 \, L }{ 2g } 
+ \Delta {\cal E}^{(0)} 
= - \frac{ \mu^2 }{ 2g } (L - z_e) 
\label{eq:scaling-gd-state-energy-flat}
\end{equation}
similar to Eq.\,(\ref{eq:def-linear-density-asymptote}) for the total particle number.
From the exact solution at a hard wall, we get from Eq.\,(\ref{eq:split-total-energy-0}): $z_e = 4/(3 \kappa)$ shown in the inset of the Figure.
For other flat-bottom potentials, results for $z_e$ are given in Table~\ref{t:ze-0-surface-energy}.
A similar analysis can be made for the particle number, whose values are given in the Table (see also Appendix~\ref{a:boundary-corrections}).

In the linear potential, the bulk density increases linearly with $z$, see Eq.\,(\ref{eq:Thomas-Fermi-density-1}).
For a proper definition of the surface energy, we have to use an expansion beyond the Thomas-Fermi approximation of Eq.\,(\ref{eq:Thomas-Fermi-density-1}).
This has been provided in Refs.\:\onlinecite{Lundh97, Margetis00}
\begin{equation}
z \gg \ell: \quad
- \frac{g}{2} \phi^4(z) = - \frac{(F z)^2}{2g} 
+ \frac{ \hbar^2 F }{ 8 g M z } + {\cal O}( 1/z^4 )
\label{eq:post-TF-density-linear}
\end{equation}
Here we use for convenience physical units, $\ell$ being the length scale in the linear potential of Table\:\ref{t:units}.
To define the surface energy $\Delta {\cal E}^{(0)}$, we replace the subtraction under the integral~(\ref{eq:split-total-energy-0}) by the leading term $(F z)^2/g \, \Theta(z)$.
The next-to-leading order gives a logarithmic dependence on the system size $L$. 
A natural parametrization for the surface energy in Eq.\,(\ref{eq:split-total-energy-0}) is thus
\begin{align}
\mbox{linear:} \quad
{\cal E}^{(0)}(L) 
& = - \frac{ F^2 L^3 }{ 6 g } 
+ \frac{ \hbar^2 F }{ 8 g M } \log \frac{L}{\ell}
+ \Delta {\cal E}^{(0)} 
\nonumber
\\
\Delta {\cal E}^{(0)}  &= 
- \frac{ \hbar^2 F }{ 8 g M } \log \frac{z_e}{\ell}
\label{eq:scaling-gd-state-energy-linear}
\end{align}
The length $z_e$ describes the contribution of the border region.
Numerically, we find $z_e \approx 1.280\,\ell$, while the pre\-factor is $\hbar^2 F / 8 g M \approx 36\,{\rm nK}$, taking the standard gravitational acceleration for $F/M$.
A logarithmic boundary correction has been identified earlier for a  harmonic potential and attributed to the kinetic energy of the condensate mode in the boundary region.\cite{Dalfovo96, Fetter98a}
Indeed, the logarithmic scaling appears if we take $(\hbar^2 / 2M) ({\rm d}\phi/{\rm d}z)^2 \sim \hbar^2 F / (8Mg\, z)$ as a proxy for the kinetic energy density and integrate over the Thomas-Fermi tail $z = z_e \ldots L$.

The first term of Eq.\,(\ref{eq:gd-state-energy-density}) is proportional to the condensate density and arises in the second order in fluctuations due the repulsive back action of particles in non-condensate modes.\cite{Mora03}
It has been discussed in Ref.\:\onlinecite{Henkel25b} and vanishes in the thermodynamic limit $L \to \infty$ at fixed density $|\phi(z)|^2$ because the prefactor $\mu'_0$ is inversely proportional to the extensive number  $N_0$ of condensate particles. 
We focus in the following on the contribution due to elementary excitations.

\subsubsection{Elementary excitations}

The second term in Eq.\,(\ref{eq:gd-state-energy-density}) that appears even at zero temperature requires the evaluation of the ``hole'' wave functions $v(z; E)$. 
Using the techniques of Refs.\:\onlinecite{Diallo15a, Henkel25c}, we solve the Bogoliubov equations numerically. 
The continuous energy spectrum is scanned with Robin boundary conditions at the right end $z = L$ of the computational domain, interpolating in this way between the sequence of eigenvalues set by Dirichlet and Neumann boundary conditions.
The results for potentials with a flat bottom have been given in Refs.\:\onlinecite{Henkel25b}, and we briefly outline them here for comparison.

\begin{figure}[htbp]
\centerline{%
	\begin{minipage}[t]{0.5\textwidth}
	\mbox{}\\
    \includegraphics[scale=0.48]{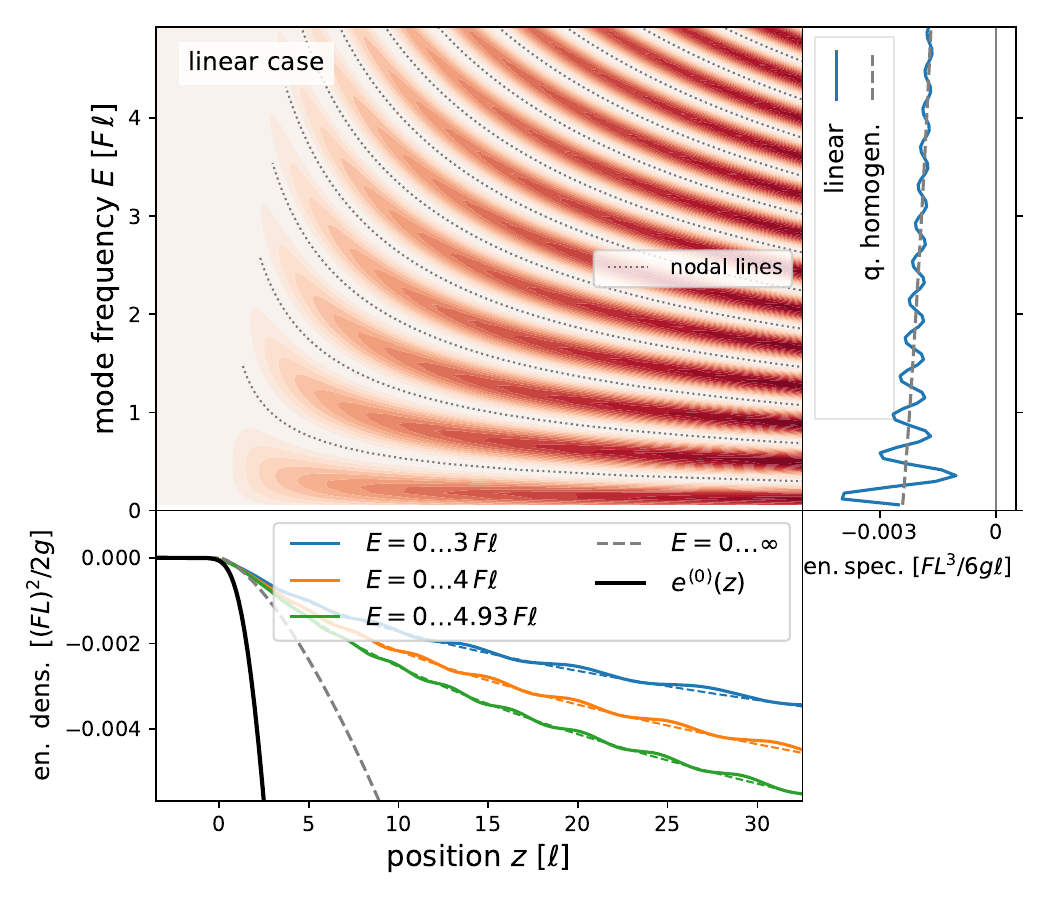}
    \end{minipage}
}
\caption[]{%
Spectral representation of the ground-state (zero temperature) energy density $-E \, v^2(z; E)$ in the border region of an inhomogeneous Bose gas confined by a linear potential (turning point at $z = 0$).
Contributions from non-condensate excitations only.
In the central contour plot, dotted are the nodal lines expected from the asymptotic form~(\ref{eq:uv-normalisation-flat-bottom}, \ref{eq:Bessel-asymptote-linear}) with $\delta = 0$.
The bottom and right panels give the energy density spectrum integrated along the energy or position axis, as indicated in the legends.
The spatial upper limit is $L = 32.5\,\ell$.
The results deep in the bulk (where mode functions 
are spatially averaged, see Eq.\,(\ref{eq:energy-density-3})) are plotted in dashed.
}
\label{fig:gd-energy-density-spectrum}
\end{figure}

For the linear potential (no flat bottom), the resulting energy density $E\,|v(z; E)|^2$ is plotted in the position-energy plane in Fig.\:\ref{fig:gd-energy-density-spectrum}.
One notes nodal lines due to the standing-wave character of $v(z; E)$ at the turning point.
These are well reproduced by the zeroes of the Bessel function $J_0(E\sqrt{2z})$ from Eqs.\,(\ref{eq:uv-normalisation-flat-bottom}, \ref{eq:Bessel-asymptote-linear}), taking a phase shift $\delta = 0$, consistent with Ref.\:\onlinecite{Henkel25c}.
There is practically no contribution from $z < 0$ because the Bogoliubov ``hole mode'' $v(z; E)$ only appears where the condensate density is nonzero, see Eqs.\,(\ref{eq:BdG-1}).
There is no infrared divergence due to the factor $E$ under the integral~(\ref{eq:gd-state-energy-density}), but the convergence at high energies is relatively slow, see also Eq.\,(\ref{eq:energy-density-asymptote-linear}) below.
As a function of position, the quantum energy density increases because the condensate pendant does the same, see Eq.\,(\ref{eq:energy-density-asymptote-linear}) below.

The panels to the right and bottom of Fig.\:\ref{fig:gd-energy-density-spectrum} illustrate the energy density integrated either along the energy or the position axis. 
The oscillatory features in the bottom panel as a function of position are artefacts. 
Technically, they arise for the same reason as Friedel oscillations in the density of a Fermi gas whose energy distribution shows a sharp edge at zero temperature.
In the Bose case, the energy dependence is smooth, however.

In a flat-bottom potential, the asymptotic form~(\ref{eq:uv-normalisation-flat-bottom}) of the hole mode $v(z; E)$ allows for an analytic evaluation of the energy integral (see Refs.\:\onlinecite{Mora03, Henkel17b}). 
We average over the spatial oscillations (denoted by the overline) and get
\begin{equation}
\int_0^E\!\frac{ {\rm d}E }{ \pi } \, E \, \overline{v^2(z; E)}
= 
\frac{ \mu \kappa }{ 2\pi } 
\Big( 
	\tfrac43 
	- {\rm e}^{-t/2} - \frac{ {\rm e}^{-3t/2}  }{ 3 }
\Big)
\label{eq:homogeneous-energy-density}
\end{equation}
Here, $\kappa$ is the inverse healing length (Tab.\:\ref{t:units}), and the integration variable $t$ arises from the substitution $E = \mu \sinh t$.
Taking the integration to infinite energy and adopting the local density approximation $\mu \mapsto \mu - V(z)$, we find for the fluctuation energy density the scaling  
\begin{equation}
e^{(2)}(z) \approx - 
\frac{2 \sqrt{M}}{3\pi\hbar}
(\mu - V(z))^{3/2}
\label{eq:scaling-energy-density-in-LDA}
\end{equation}
provided $V(z) \le \mu$.
The unusual scaling with $\hbar$ is a reminder that quantum fluctuations in a Bose gas arise from particle interactions and second quantization.

A similar calculation can be performed for the linear potential.
We focus on the asymptote $E \sqrt{2z} \to \infty$ where the mode functions Eq.\,(\ref{eq:Bessel-asymptote-linear}) simplify using the large-argument expansion of the Bessel functions.\cite{DLMF}
Written in natural units $F = \ell = 1$, 
\begin{align}
E \, \overline{v^2(z; E)} & \approx 
\sqrt{2z} \, \sin^2 [ \tfrac12 \theta(z) ] \,
\overline{\sin^2( E \sqrt{2z} + \pi/4 )}
\nonumber\\
& \approx
\displaystyle
\frac{ z^2 \sqrt{2z} }{ 4 C(z) (E + C(z)) }
\label{eq:energy-density-3}
\end{align}
where Eq.\,(\ref{eq:def-theta-linear}) has been used for $\theta(z)$,
and $C^2(z) = E^2 + z^2$.
The integral over the position $z$ leads to an elliptic integral (not displayed
for brevity) that also agrees well with the numerical data 
(Fig.\:\ref{fig:gd-energy-density-spectrum}, right panel, ``quasi-homogeneous'' system), 
except for the oscillations.
The energy integral can be performed with the substitution $E = z \sinh t$,
and yields
\begin{equation}
\mbox{linear:}\quad
\int_0^E\!\frac{ {\rm d}E }{ \pi } \, E \, \overline{v^2(z; E)}
\approx
\frac{ F z }{ 4 \pi \, \ell} \sqrt{ \frac{ 2 z }{ \ell } }
\left( 1 - {\rm e}^{ - t} \right) 
\label{eq:energy-density-asymptote-linear}
\end{equation}
where we have re-instated physical units.
Note the slow convergence ($\sim 1/E$) in energy, with a different exponent than in Eq.\,(\ref{eq:homogeneous-energy-density}), however.
Using the definition of $\ell$ in Table~\ref{t:units}, we recover the scaling of Eq.\,(\ref{eq:scaling-energy-density-in-LDA}) with the local chemical potential $\mu - V(z) = F z$ and with $\hbar$.
The result~(\ref{eq:energy-density-asymptote-linear}) aligns with the numerics for a finite upper limit (bottom panel, thin dashed lines).

We finally note that the fluctuation corrections are everywhere small compared to the ground state energy density and its integrals (up to the length $L$ shown in Fig.\:\ref{fig:gd-energy-density-spectrum}).
The latter, which is proportional to $z^2$ in the Thomas-Fermi regime, is plotted as a thick black line in the bottom panel.
Relative to the energy correction due to Bogoliubov excitations, this is a factor $\approx 21\,(z/\ell)^{1/2}$ larger, using the numbers of Tab.\:\ref{t:units}.

\subsection{Finite temperature}
\label{s:energy-density-excitations}

The average of the second-order energy density~(\ref{eq:expand-H2-step-1}) in thermal equilibrium is found using Eqs.\,(\ref{eq:P-square-average}, \ref{eq:b-bdag-average})
\begin{align}
\langle
\hat{K}^{(2)}(z)
\rangle_T
= \ & T \, \phi_a \phi
\label{eq:average-K2}
\\
& {} 
+
\int_0\!\frac{{\rm d}E}{\pi} 
\, E
\, \bose(E) 
\left(
|u(z; E)|^2 
-
|v(z; E)|^2
\right)
\nonumber
\end{align}
where we recall that $\bose(E)$ is the Bose distribution.

The first term, depending only on the condensate, has been interpreted by Mora and Castin in the context of particle number fluctuations in the grand-canonical ensemble.\cite{Mora03}
Indeed, recalling that the operator $\hat{P}$ gives the fluctuation in the number of condensed particles, the first term in Eq.\,(\ref{eq:average-K2}) is given by
\begin{equation}
T \phi_a \phi = \mu'_0 \frac{\langle \hat{P}^2 \rangle}{2}
\frac{ \partial |\phi|^2 }{ \partial N_0 }
\label{eq:average-K2-condensate}
\end{equation}
Up to a global factor, this coincides with the density res\-ponse $\partial |\phi|^2 / \partial \mu$.
The latter is plotted in Fig.\:\ref{fig:sketches} (solid curves) and becomes a constant $1/g$ for both types of confining potentials, as predicted by Eq.\,(\ref{eq:Thomas-Fermi-density-1}).
The contribution~(\ref{eq:average-K2-condensate}) is vanishing in the thermodynamic limit $N_0 \to \infty$.

\begin{figure}[htbp]
   \centering
   \hspace*{-02mm}%
   \includegraphics[height=0.4\columnwidth]{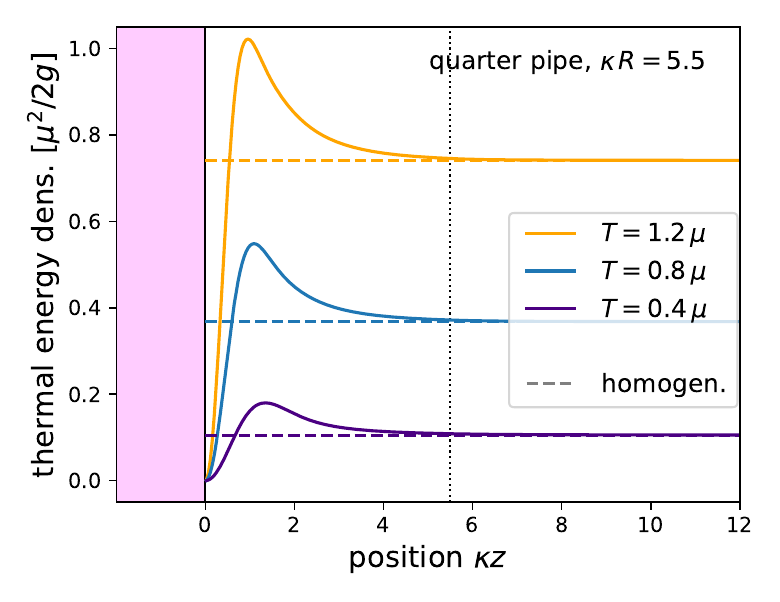}
   \hspace*{-02mm}%
   \includegraphics[height=0.4\columnwidth]{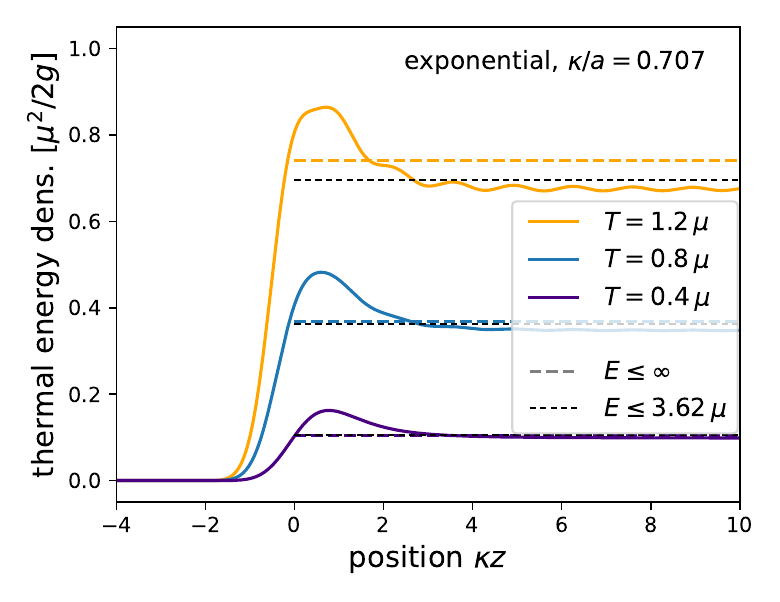}
   \hspace*{-02mm}%
\caption[]{Thermal energy density, contribution from quasi-particles. 
(\emph{left}) hard barrier with quarter-pipe potential,
(\emph{right}) exponential (soft) potential.
Solid (black thin dashed) lines: numerical (analytical) calculation up to a finite energy;
colored thick dashed lines: analytical prediction at infinite energy cutoff for homogeneous system. 
Data normalised to the bulk condensate energy density at $T = 0$.
}
   \label{fig:thermal-e-dens-exp}
\end{figure}

The second term in Eq.\,(\ref{eq:average-K2}) is plotted in Fig.\:\ref{fig:thermal-e-dens-exp}, the integration over energy being performed.
Here confinements by the quarter-pipe and the exponential barrier are considered.
Note the peak near the turning point where non-condensed particles are filling the potential well formed around $z = 0$ by the confinement potential and the repulsive mean-field potential.
The horizontal lines illustrate the results for a homogeneous system obtained as in Eq.\,(\ref{eq:homogeneous-energy-density}) above, the thin dashed lines being evaluated for the same energy cutoff as the numerical data. 
(We attribute slight discrepancies with the numerics to an insufficiently fine sampling at low energy values.)
The oscillatory Friedel artefact visible at the highest temperature is due to the sharp energy cutoff in the numerical data (right panel).

\section{Particle density}
\label{s:depletion-thermal}

\subsection{General features}

As a second application, we consider the particle density $n(z; T) $ in thermal equilibrium, usually split into condensate $n_c(z; T)$ and excitation $n_x(z; T)$ parts.
We follow again the thermodynamic theory of \citet{Mora03} whose equilibrium statistics is governed by the Hamiltonian~(\ref{eq:H2}).
Technically, the calculations are more involved because the fluctuating Bose field is described by the operators $\delta \hat n$ and $\delta \hat \varphi$ for density and phase variations.
A non-condensate density $\langle \delta n \rangle \ne 0$ appears only in the third order in fluctuations.
A few details are sketched in Appendix~\ref{a:n-prime-3rd-order}.
The two points discussed here address (i) an alternative view on the non-condensate density based on Hartree-Fock-Bogoliubov theory [Eq.\,(\ref{eq:HFB-for-true-phi}) below] and (ii) its infrared behavior in the limit of a continuous spectrum.

\subsection{Hartree-Fock-Bogoliubov theory}
\label{s:HFB-vs-Mora03}

This theory aims at a self-consistent model for the Bose gas with two components, a ``true condensate'' $\phi_c$ and ``excited particles'' with density $n_x$.
By construction, the total density is $n = |\phi_c|^2 + n_x$. 
An ``anomalous density'' $m_x = \langle \hat{B}\hat{B} \rangle$ appears because the Bose fluctuation field 
\begin{equation}
\hat{B}(z) = 
\int_0\!\frac{ {\rm d}E }{ \mbox{\small${\sqrt{\pi}}$} } 
\left[ u(z; E)\, \hat{b}(E) + v(z; E)\, \hat{b}^\dag(E)\right]
\label{eq:uv-mode-expansion}
\end{equation}
mixes creation and annihilation operators (Bogoliubov transformation).
We insert $\Psi = \phi_c + \hat{B}$ into the Hamiltonian $K$ [Eq.\,(\ref{eq:gcan-Hamiltonian})] and work out averages using Wick's theorem (i.e., adopting approximately Gaussian statistics for $\hat{B}$).
The variation with respect to $\phi_c$ then yields the Hartree-Fock-Bogoliubov equation\cite{PitaevskiiStringari, ProukakisBook, Griffin96, Fedichev98, Rusch99, Giorgini00}
\begin{equation}
- \frac{ {\rm d}^2\phi_c }{ {\rm d}z^2 } + V \,\phi_c 
+ g \left( |\phi_c|^2 + 2 n_x \right) \phi_c + g m_x \, \phi^*_c 
= \mu \,\phi_c
\label{eq:HFB-for-true-phi}
\end{equation}
The factor $2$ in front of $n_x = \langle \hat{B}^\dag \hat{B} \rangle$ is in the spirit of Hartree-Fock theory (direct and exchange terms).\cite{CohenTannoudji_vol3}
It is absent for the condensate which is assumed to be in a 
coherent state. 

Assume now that the non-condensate densities $n_x$, $m_x$ are small compared to $|\phi_c|^2$, and expand $\phi_c = \phi - \eta_B$ to first order around the zeroth-order condensate $\phi$, that solves the Gross-Pitaevskii equation~(\ref{eq:GPe-1}).
This yields
\begin{equation}
H_3 \, \eta_B
=
g \phi \left(2 n_x + m_x \right)
\label{eq:HFB-for-etaB}
\end{equation}
where the collective Schrödinger-type operator $H_3 = H_3(\phi)$ was defined in Eq.\,(\ref{eq:H3-chi}).
In the perturbative scheme of \citet{Mora03}, the Bogoliubov modes $u$, $v$ in Eq.\,(\ref{eq:uv-mode-expansion}) are constructed from the real field $\phi$ so that the anomalous average $m_x$ is real, too.
With this proviso, Eq.\,(\ref{eq:HFB-for-etaB}) is equivalent to Eq.\,(\ref{eq:Mora03-for-etaB}) in Appendix~\ref{a:n-prime-3rd-order}.

Keeping in mind the expansion around $\phi$, the total density is
\begin{eqnarray}
|\phi_c|^2 + n_x 
&\approx& 
|\phi|^2 - 2\mathop{\rm Re} \phi \,\eta_B + n_x 
\nonumber\\
&=& \phi^2 - 2 \phi \,\eta_B + \langle \hat{B}^\dag \hat{B} \rangle
\label{eq:subtracted-density}
\end{eqnarray}
Relative to the zeroth order condensate density $\phi^2$, we thus
recover the $B$-dependent terms of Eq.\,(\ref{eq:65-Mora03-2}).
--
The theory of \citet{Mora03} provides corrections to this picture that arise from grand-canonical fluctuations in the particle number and from the commutator of $\hat{B}$ involving the condensate mode (see Eq.\,(\ref{eq:fluctuation-operator})).
These are discussed in Appendix~\ref{a:n-prime-3rd-order}, Fig.\,\ref{fig:n-prime-GC}, but turn out to be sub-leading in the thermodynamic limit.

\subsection{Infrared-regular subtraction scheme}
\label{s:IR-regular-subtraction}

The quasiparticle spectrum of the Bose gas in a half-open potential is gapless, entailing well-known infrared divergences.\cite{PitaevskiiStringari, ProukakisBook}
Not only does the Bose function $\bar{n}(E)$ blow up as $E \to 0$, but also the normalisation of the mode functions $u$ and $v$ with $v(z;E) \to -u(z;E)$ [see Eqs.\,(\ref{eq:u-v-orthogonality}, \ref{eq:uv-normalisation-flat-bottom}--\ref{eq:def-jy-flat-bottom})].
Expressions like $\langle \hat{B}^\dag \hat{B} \rangle$ thus do not make sense.\cite{Henkel17b} 
We show here that the total density is regular, however, provided the two terms that are added to/subtracted from $\phi^2$ in Eq.\,(\ref{eq:subtracted-density}) are evaluated mode by mode.
We do this by working first at fixed energy and integrating at the very end.
Due to operator ordering, it turns out that at zero temperature, a nonzero density correction emerges that is usually called depletion.\cite{PitaevskiiStringari}
The procedure illustrates that due to mutual interactions, the distinction between condensate and non-condensate densities is blurred.
We also recall that the energy density (Sec.\:\ref{s:energy-density-excitations}) does not suffer from infrared divergences because of the additional factor $E$ in Eq.\,(\ref{eq:average-K2}).

We introduce the spectral representation for the source term of the inhomogeneous Schrödinger equation~(\ref{eq:HFB-for-etaB}):
\begin{align}
g \phi \left(2 n_x + m_x \right) &=
\int_0\!\frac{{\rm d}E}{\pi} \, S_B(z; E)
\nonumber\\
S_B(z; E) &= 
2 g \phi(z) \,
\bose(E) \left[ u^2(z) + v^2(z) + u(z) v(z)
\right]
\nonumber\\
& \hphantom{=\ } {} +
g \phi(z) \left( u(z) + 2 v(z) \right) v(z)
\label{eq:H3-for-chi-at-energy-E}
\end{align}
This is found by using the mode expansion~(\ref{eq:uv-mode-expansion}) and the thermal averages~(\ref{eq:b-bdag-average}) for the continuum operators $\hat{b}(E)$, $\hat{b}^\dag(E)$.
By solving $H_3 \eta_B(z; E) = S_B(z; E)$, we get the spectral representation $\eta_B(z; E)$ of the condensate back-action.
Define $dn^{(2)}(z; E)$ as the spectral increment due to modes with energy $E$ to the quantity $\langle \hat{B}^\dag \hat{B} \rangle - 2 \phi \,\eta_B$.
Working again through the averages of the Bogoliubov expansion, we get
\begin{eqnarray}
&& {\rm d} n^{(2)}(z; E) = 
\label{eq:subtraction-per-mode}
\\
&& \frac{{\rm d}E}{\pi}
\Big\{
\bose(E) \left[ u^2(z) + v^2(z) \right]
+
v^2(z)
- 2 \phi(z) \, \eta_B(z; E) 
\Big\}
\nonumber
\end{eqnarray}
where the energy dependences of $u, v$ have been suppressed for simplicity.

Numerically, the inhomogeneous Schrödinger equation is solved on a spatial grid using a finite-difference scheme for $H_3$.
One has to deal, however, with the fact that the source term in Eq.\,(\ref{eq:HFB-for-etaB}) is spatially extended rather than localised in the border region.
This leads to an unphysical, exponentially growing ``tail'' close to the end $z = L$ of the numerical grid.
The tail arises because the homogeneous equation $H_3 \chi = 0$ allows for an exponentially increasing $\chi$ like an unphysical ``tunnelling'' solution below the smooth potential in $H_3$.
We reduce its amplitude by extending the numerical grid beyond $z = L$ for a few units, determined by the growth rate of the tunnelling tail.
At the end of the extended grid, a logarithmic derivative boundary condition is applied based on the Thomas-Fermi approximation
\begin{equation}
\eta_B(z; E) \approx \frac{ S_B(z; E) }{ 2 g \phi^2(z) }
\label{eq:TF-like-solution-for-chi-B}
\end{equation}
We recall that this solution neglects the second derivative in $H_3$  and uses for its potential the form $V_3(z) \approx 2 g \phi^2(z)$ [see Eqs.\,(\ref{eq:Thomas-Fermi-density-1}, \ref{eq:H3-chi})].

The approximate solution~(\ref{eq:TF-like-solution-for-chi-B})
clarifies the link between the extended Bogoliubov theory\cite{Mora03}
and other formulations that have been based more or less explicitly
on the local density approximation.\cite{AlKhawaja02b, Henkel17b}
Inserting Eq.\,(\ref{eq:TF-like-solution-for-chi-B}) into the
subtraction scheme of Eq.\,(\ref{eq:subtraction-per-mode}), we get
\begin{equation}
{\rm d} n^{(2)}(z; E) \approx
{-} \frac{{\rm d}E}{\pi} 
\Big\{
\bose \,
f(z)^2 
+
f(z) 
\, v(z)
\Big\}
\label{eq:TF-density2-after-subtraction}
\end{equation}
where $f(z; E) = u(z; E) + v(z; E)$ corresponds to quasi-particle density fluctuations\cite{Diallo15a, ProukakisBook} (see Fig.\:\ref{fig:density-drape} below).
Up to a sign, this essentially agrees with the regularised non-condensate density of modified Popov theory.\cite{Henkel17b}
While the temperature-dependent part (proportional to the Bose distribution $\bose$) is negative, the quantum contribution is positive.
(Recall Eq.\,(\ref{eq:uv-normalisation-flat-bottom}) for our sign convention of the $u$- and $v$-modes.)
We have checked that the result~(\ref{eq:TF-density2-after-subtraction}) is a good approximation deep in the condensate.
Close to the border, however, density gradients become significant, and only the scheme of Eqs.\,(\ref{eq:HFB-for-etaB}, \ref{eq:subtracted-density}) can be used beyond the local-density approximation, as we shall see now.

\subsection{Results}

\begin{figure}[tbhp]
   \centering
   \includegraphics[width=0.95\columnwidth]{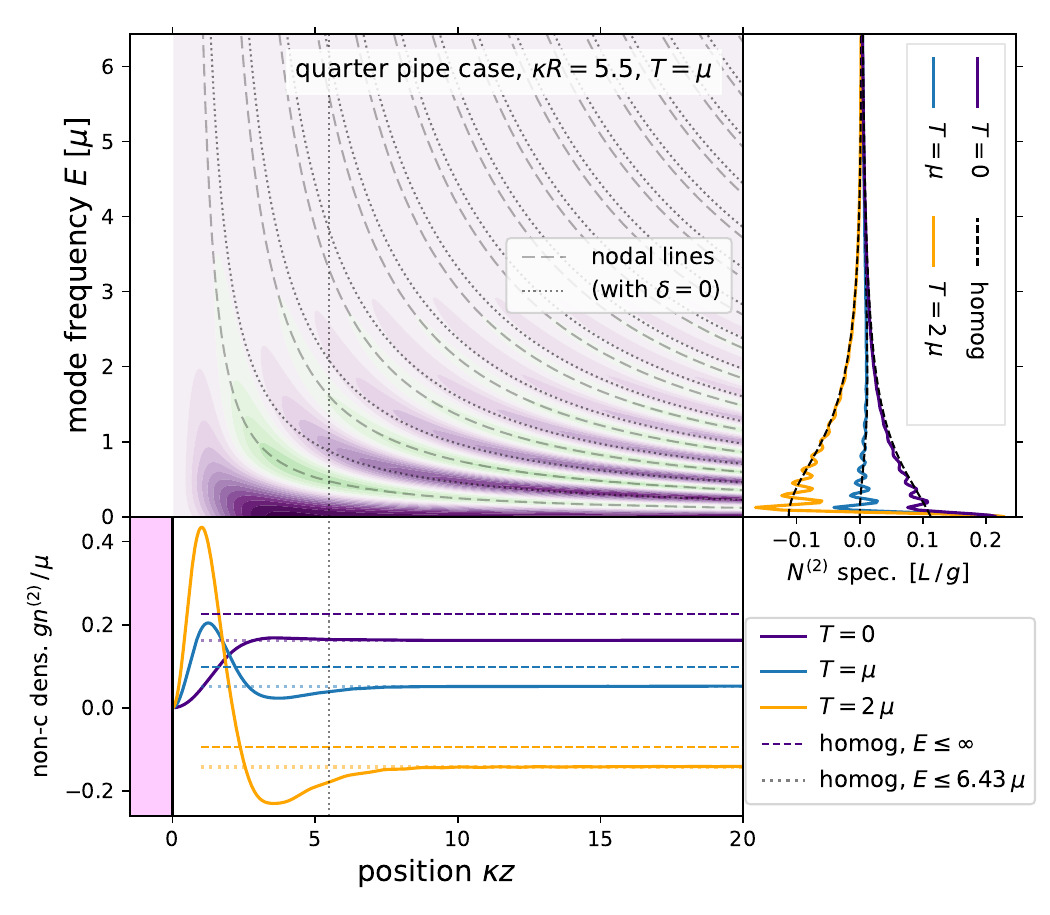}
\caption[]{%
Non-condensate density near the turning point in the quarter pipe potential (Dirichlet boundary condition at $z = 0$ combined with a rounded potential between $z = 0$ and $z = R$).
Central panel: space- and energy-resolved contribution ${\rm d}n^{(2)}(z; E)/{\rm d}E$ of Eq.\,(\ref{eq:subtraction-per-mode}). 
Purple (green) color: positive (negative) values, respectively.
In dashed and dotted, the nodal lines expected from the large-distance asymptote of the Bogoliubov wave functions.
Right panel: spatially integrated density from $z = 0 \ldots L \approx 20/\kappa$ vs.\ mode energy; in dashed, predictions for a homogeneous system of size $L$ where oscillations of mode functions are averaged out.
Bottom panel: energy-integrated density. 
For the solid lines, integrations are performed over the energy range shown in the central plot.
Set of dashed (dotted) lines for $z > 0$: integration to $E = \infty$ ($E = 6.43\,\mu$) for a homogeneous system. 
The integrated quantities are normalised to those for a homogeneous system with density $\mu / g$ of size $L$.
}
   \label{fig:square-density-quarter-pipe}
\end{figure}

\begin{figure}[tbhp]
   \centering
   \includegraphics[width=0.95\columnwidth]{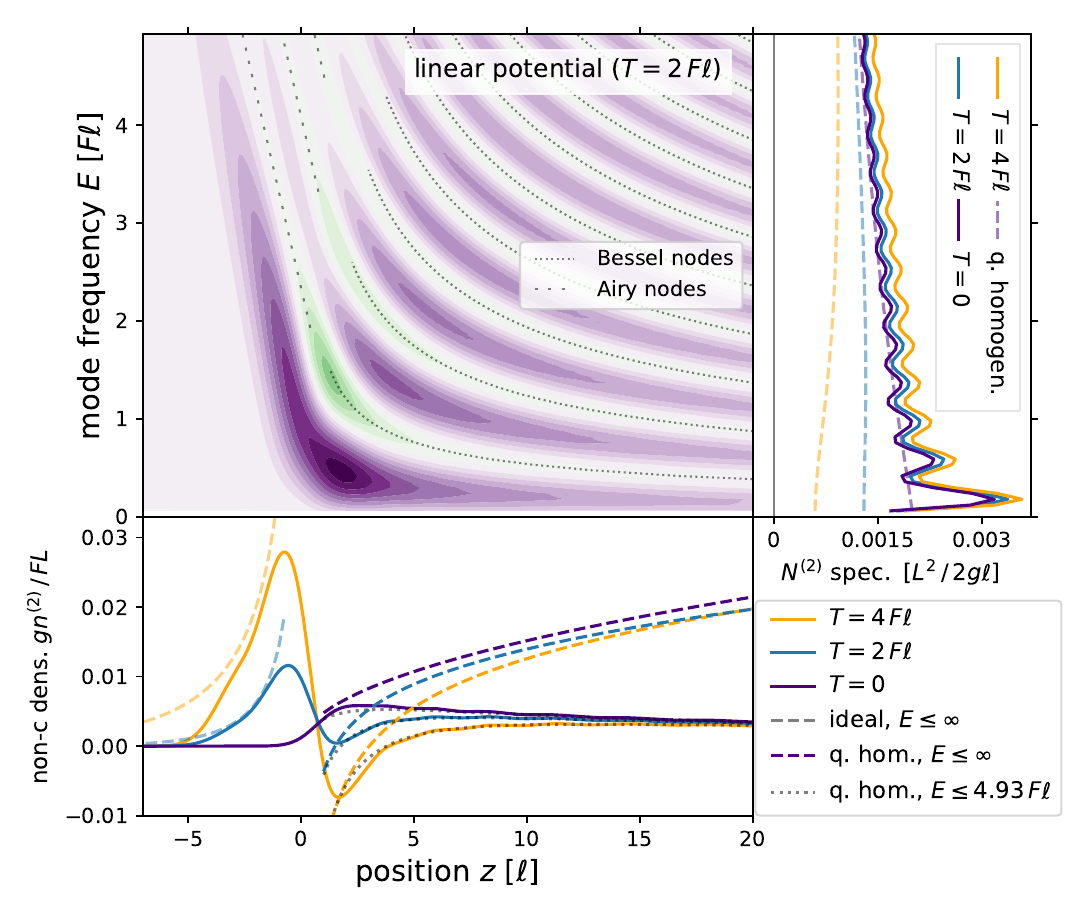}
\caption[]{%
Similar to Fig.\:\ref{fig:square-density-quarter-pipe}, but for the linear potential.
Central panel: in dotted, the nodal lines expected from the low-density (Airy) and high-density (Bessel) [Eqs.\,(\ref{eq:uv-normalisation-flat-bottom}, \ref{eq:Bessel-asymptote-linear}) with $\delta = 0$] asymptotes of the Bogoliubov wave functions.
Positions are normalised to the natural length scale $\ell$ in the linear potential (Table~\ref{t:units}).
Right panel: spatially integrated density vs.\ mode energy; for the quasi-homogeneous system (dashed), the oscillations of the mode functions are averaged out.
The integration is performed over the range $x = -12\,\ell \ldots L = 32.5\,\ell$, but only a smaller range is displayed for better visibility.
Bottom panel: energy-integrated density.
Solid lines: over the displayed range; 
dashed (dotted) lines for $z > 0$: integration up to $E = \infty$ ($E = 4.93\,F \ell$) for a quasi-homogeneous system. 
Faded dashed lines for $z < 0$: ideal Bose gas with the local chemical potential $\mu(z) = \mu - V(z)$.
The integrated quantities are normalised to the values corresponding to the Thomas-Fermi condensate evaluated at $z = L$ (bottom panel) or integrated from $z = 0$ to $L$ (right panel).
}
   \label{fig:square-density-linear}
\end{figure}

Typical results of this calculation are shown in Figs.\:\ref{fig:square-density-quarter-pipe} and~\ref{fig:square-density-linear} for the quarter pipe and linear potentials.
The central contour plot shows ${\rm d} n^{(2)}(z; E)/{\rm d}E$ for a given temperature, while the panels to the right and bottom show integrated quantities, using the ranges of positions and energies displayed in the Figure.
Nodal lines appear similarly to Fig.\:\ref{fig:gd-energy-density-spectrum} that reveal the energy-dependent phase shift $\delta$ in Eq.\,(\ref{eq:uv-normalisation-flat-bottom}).\cite{Henkel25c}
The limiting values for $E \to 0$ are finite.
To interpret the general behaviour, we recall that ${\rm d} n^{(2)}(z; E)$ gives the density change relative to the condensate density $\phi^2(z)$.
In the region $z \lesssim 2/\kappa$ of small condensate density, additional density is created by thermally populated Bogoliubov excitations.
Further away from the turning point, the total density is \emph{reduced}, however, because the subtraction $-2 \phi(z) \eta_B(z)$ (recall the back action of the additional mean-field potential) over-compensates for the Bogoliubov density $\langle B^\dag(z) B(z) \rangle$ in Eq.\,(\ref{eq:subtracted-density}).
This leads to a pronounced dipolar feature in the excess density visible in Fig.\:\ref{fig:square-density-quarter-pipe}, bottom panel.
The dashed and dotted lines plotted here (and in the right panel) correspond to the predictions of Eq.\,(\ref{eq:TF-density2-after-subtraction}) evaluated for a homogeneous system (constant density, oscillatory wave functions averaged out, similar to the expressions discussed in Sec.\,\ref{s:energy-density}).
They illustrate in particular the convergence when all energies are taken into account: obviously the quantum depletion part is not cut off by the Bose function.

Despite the oscillatory structure of the density spectrum ${\rm d} n^{(2)}(z; E)$, there is no trace of Friedel oscillations once the energy integration is carried out, apart from the broad dipolar structure noted above.
Note that for the chosen temperatures, the non-condensate density reaches a moderate fraction of the condensate density (between $+40\%$ and ${-}20\%$ for $T = 2\,\mu$): 
this announces the breakdown of the Bogoliubov approximation.
The results for the hard-wall and exponential potentials are similar (not shown).

In the linear potential (Fig.\:\ref{fig:square-density-linear}), the inhomogeneous condensate density somewhat changes the picture.
Because the potential is soft, nonzero temperature adds non-condensate particles to the region $z \lesssim 0$ beyond the turning point (thick purple patch).
There, the Bogoliubov modes are dominated by their $u(z;E)$-part that behaves like an Airy function, since outside the bulk condensate, only the potential energy is relevant. 
This can be verified from the nodal lines with wide dotting in the main panel.
Upon integration over the energy, the density in this region depends on position and temperature roughly as predicted by the ideal gas equation of state in the local-density approximation, as illustrated by the faded dashed lines in the bottom panel.
(The deviations visible for $z \lesssim -2\ell$ are due to the finite energy range behind the numerical data.)
On the other side $z \gtrsim \ell$ of the turning point and for a sufficiently large temperature, the thermal density takes over with a negative sign so that a dipolar structure appears, similar to Fig.\:\ref{fig:square-density-quarter-pipe}.
The zero-temperature (depletion) contribution becomes dominant, however, as one moves into the bulk condensate. 
In particular, when the entire energy spectrum is integrated over, the non-condensate density remains positive up to all distances from the turning point shown here (coloured dashed lines).

To obtain the quasi-homogeneous data shown in the Figure, we have averaged over the oscillations of the $f, v$ mode functions in Eq.\,(\ref{eq:TF-density2-after-subtraction}) and used their asymptotic form, similar to Eq.\,(\ref{eq:energy-density-3}).
This gives good agreement in the spatial range $z \gtrsim 0$ (dotted lines in bottom panel). 
The plot also illustrates the weak convergence of the energy integration (compare solid and dashed lines), again due to the depletion density. 
In this soft potential, the range $z \lesssim 0$ contributes significantly to the total (non-condensate) particle number (right panel in Fig.\:\ref{fig:square-density-linear}) and is not captured at all by the bulk density. 
This explains the significant differences between the numerical particle number (solid lines) and its homogeneous approximation (dashed lines).

Note finally the relatively small numbers for the non-condensate densities in Fig.\:\ref{fig:square-density-linear}. 
They arise because we normalise to the condensate density $\phi^2(L) \approx F L / g$ evaluated at the position $z = L = 32.5\, \ell$ in the linearly increasing condensate background.
For a fair comparison between Figs.\,\ref{fig:square-density-quarter-pipe}, \ref{fig:square-density-linear}, take into account that the energy units in the linear and flat-bottom potentials are roughly $F \ell : \mu  \sim 1:3$.

\section{Conclusion}

A detailed analysis of the spatial distribution of particles and energy near the border of a one-dimensional, ultracold gas of interacting Bosons has been given.
Although at finite temperature, there is no proper Bose-Einstein condensate (large phase fluctuations in one dimension), the concept of a quasi-condensate with reduced density fluctuations provides a convenient starting point.\cite{Mora03,ProukakisBook}
The border of this system is characterized by a gas density that is smoothly dropping, connecting a dense phase with a dilute ideal gas.
The elementary (Bogoliubov) excitations in the border region require computational tools beyond the local-density approximation precisely because of steep density gradients.
A generic conclusion that can be drawn from this and previous papers is that the condensate surface region has physical characteristics on its own.
Virtual particles depleting the condensate or a finite-temperature population may create local density modifications that have the opposite sign compared to a homogeneous system (e.g., Fig.\:\ref{fig:square-density-quarter-pipe}).
One may speculate whether these are stationary-state traces of the ``sound'' radiation that is dynamically generated when a dark soliton coming from the bulk is turning around near the border region, by a breakdown of integrability.\cite{Konotop08, Burger99, Jackson_2007a}

This paper concludes a series of investigations starting with a convergent scheme by \citet{Diallo15a} to solve the Bogoliubov--de Gennes equations in the continuous spectrum.
The phase shift of the corresponding standing waves is dispersive for barriers with a flat bottom, but vanishes practically for a linear confinement, it is also quite distinct from the one predicted by the local-density approximation.\cite{Henkel25c}
The spatial energy profile gives access to the surface energy of the Bose gas which amounts to the equivalent of a fraction of one particle, but again with contributions of different signs.\cite{Henkel25b}
We note that this energy (as in Fig.\:\ref{fig:gd-energy-density-spectrum}) can be converted into a spectrogram and listened to, a method called sonification.\cite{Henkel26a}
Finally, in this paper, density and non-condensate energy at finite-temperature have been analyzed, clarifying the links between perturbative and self-consistent descriptions.

\begin{figure}[htbp]
   \centering
   \includegraphics[width=0.85\columnwidth]{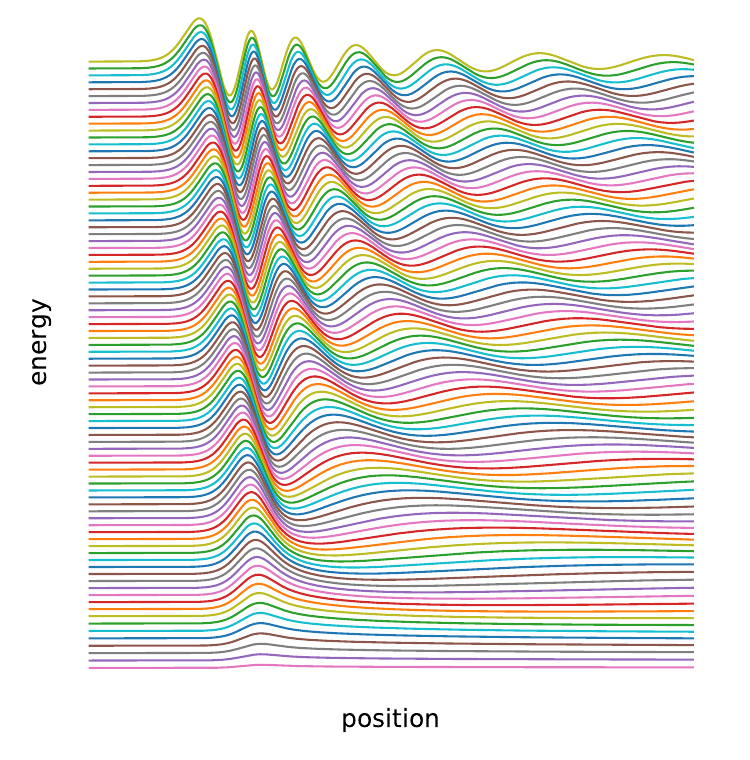}
   \caption[]{The spectrum of density modes $f(z; E)$ in a linear potential: farewell, Rudi.}
   \label{fig:density-drape}
\end{figure}

\paragraph*{Acknowledgments.}
For calculations and discussions at various stages of this work, I thank Uwe Bandelow, Abdoulaye Diallo, Enrico Reuß, Anja Seegebrecht, and the late Gabriel Barton.
--
This paper is dedicated to the memory of Rudolf (Rudi) Podgornik who inspired generations of research in finite-temperature interactions in complex systems.
The wavy spectrum of density modes $f(z; E)$ displaying in Figure~\ref{fig:density-drape} may be understood as a final greeting.

The work of C.\,H. is supported by the Deutsche Forschungsgemeinschaft (German Research Foundation) within SFB 1636, ID 510943930, Projects No. A01 and A04.

\appendix

\section{Boundary corrections to particle number}
\label{a:boundary-corrections}

\subsection{Flat-bottom potentials}
\label{a:dNc_border}

\begin{figure}[tbp] 
   \centering
   \includegraphics[width=0.85\columnwidth]{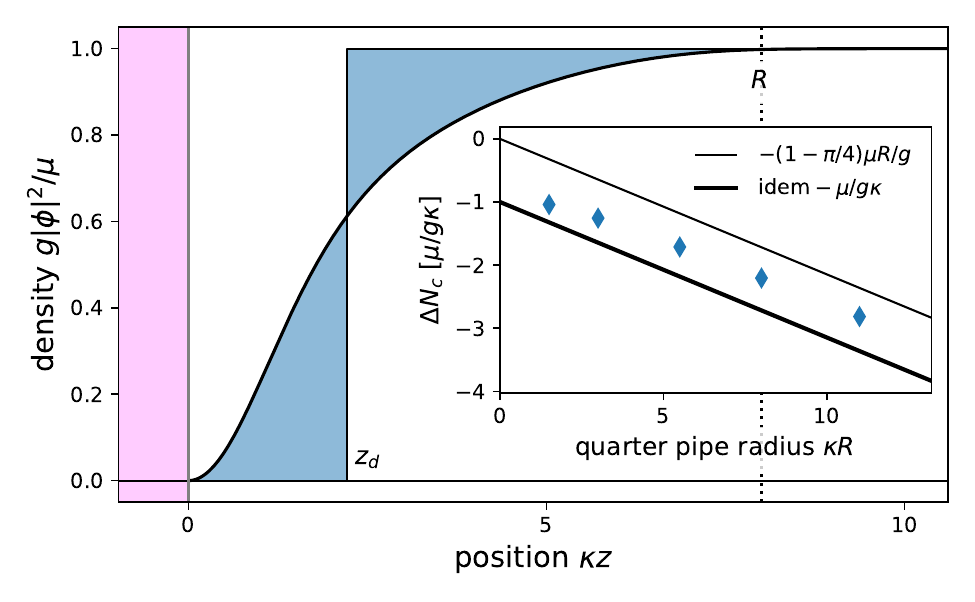}
   \caption[]{%
   The hard-wall boundary condition repels particles and leads to a depletion zone near $z = 0$ whose area (particle number $\Delta N_c$) is the same as the rectangle between $z = 0 \ldots z_d$.    
   Quarter-pipe potential with radius $\kappa R = 8$.
   Inset: missing particle number $\Delta N_c$ vs.\ $R$.
   }
   \label{fig:dNc_quarter_pipe}
\end{figure}

The density $|\phi(z)|^2$ of the condensate particles smoothly joins
the Thomas-Fermi asymptote~(\ref{eq:Thomas-Fermi-density-1}) as $z$ moves
away from the turning point $z = 0$, see 
Figs.\:\ref{fig:sketches}, \ref{fig:dNc_quarter_pipe}.
Its particle number, integrated up to a (large) reference position 
$z = L$, therefore behaves like
\begin{equation}
N(z \le L) = \int\limits_{-\infty}^{L}\!{\rm d}z \, |\phi(z)|^2 
\approx \frac{ \mu L }{ g } + \Delta N_c =
n_c ( L - z_d )
\label{eq:def-linear-density-asymptote}
\end{equation}
for flat-bottom potentials where the condensate density approaches 
the constant $n_c = \mu/g$.
(For the linear potential, see Appendix~\ref{a:dipole}.)
The missing particle number $-\Delta N_c = n_c z_d$ 
arises from the hole (notch) enforced by the boundary condition.
At a hard wall, we find explicitly
\begin{equation}
\Delta N_c = \frac{\mu}{g} \int\limits_0^{L}\!{\rm d}z 
\left(
\tanh^2(\kappa z) - 1
\right)
\approx - \frac{\mu}{g \kappa}
\label{eq:dNc-for-dark-soliton}
\end{equation}
or $z_d = 1/\kappa$. 
(We took $L \gg 1/\kappa$ in the last step.)
With the parameters of Table~\ref{t:units}, the scale factor is $\mu/g\kappa \approx 12$.
For the quarter-pipe potential, the construction is illustrated in Fig.\:\ref{fig:dNc_quarter_pipe}. 
The shaded areas outline a profile with a step at $z_d$ that gives the same number of condensate particles, similar to the condition of charge neutrality at a ``jellium'' surface.\cite{Feibelman82, Toscano15}
The inset illustrates how $\Delta N_c$ varies with the quarter-pipe radius $R$.
The lines give upper and lower limits that are found either by integrating over the Thomas-Fermi approximation (the difference between a rectangle and a quarter ellipse) or by adding to this result the ``missing hole'' of Eq.\,(\ref{eq:dNc-for-dark-soliton}).
The scaling is roughly $\Delta N_c \propto - n_c R$.
A qualitatively similar behaviour is found for the exponential potential where $\Delta N_c \propto - n_c / a$.

\subsection{Exact density balance in the linear potential}
\label{a:dipole}

\begin{figure}[htbp] 
   \centering
   \includegraphics[width=0.75\columnwidth]{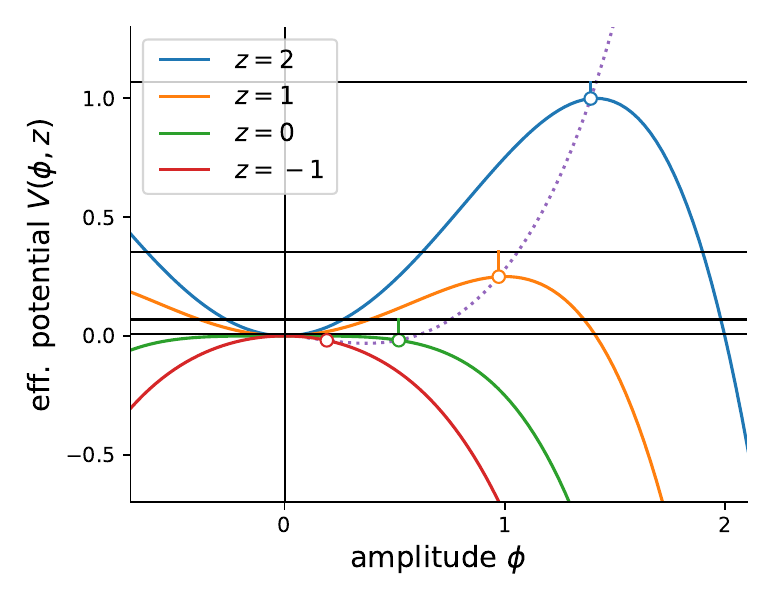} 
   \caption{%
   Potential $V(\phi; z)$ of Eq.\,(\ref{eq:GPe-Hamiltonian}) in the interpretation of $\phi( z )$ as a dynamical system. 
   We use natural units for the linear potential here.
   The dotted line represents the physical solution, white circles marking its values $\phi(z)$ at the chosen positions. 
   The horizontal lines are the corresponding instantaneous energies $H(\phi, p; z)$ with the kinetic energy becoming smaller and smaller.
   The Thomas-Fermi asymptote corresponds to the ``top of the hill''.
   }
   \label{fig:potential}
\end{figure}

The linear potential where the Gross-Pitaevskii wave function
$\phi(z)$ is given by the second Painlev\'e transcendent
\cite{Fetter98a},
has the property that $\Delta N_c = 0$ exactly.
Indeed, we are going to prove 
\begin{eqnarray}
&& N(z \le 0) + \delta N(z \ge 0) = 0
	\label{eq:total-density-conserved}
\\
&& \mbox{with}\quad
\delta N(z \ge 0) 
= \int\limits_{0}^{\infty}\!{\rm d}z 
\left(
|\phi(z)|^2 - F z / g 
\right)
\,,
\nonumber
\end{eqnarray}
where the Thomas-Fermi density $|\phi(z)|^2 = F z/g$ is subtracted
to make the integral converge.
This identity also appears as integral $I$ in Ref.\:\onlinecite{Fetter98a}, Appendix~A.
The key idea is to translate the nonlinear Schrödinger equation~(\ref{eq:GPe-1}) into a dynamical system: 
read $z$ as a ``time coordinate'', $\phi( z )$ as a ``displacement'' and $p( z ) = {\rm d}\phi / {\rm d}z$ as a ``momentum''.\cite{Miron25}
The wave equation is then equivalent to the mechanical problem
\begin{equation}
\frac{ {\rm d}\phi }{ {\rm d} z } =
\frac{ \partial H }{ \partial p }
\,,\qquad
\frac{ {\rm d}p }{ {\rm d} z } =
- \frac{ \partial H }{ \partial \phi }
\label{eq:GPe-in-canonical-form}
\end{equation}
with the Hamilton function
\begin{eqnarray}
H( \phi, p; z) &=& 
\frac{ p^2 }{ 2 } + V( \phi; z )
\,,
\label{eq:GPe-Hamiltonian}\\
V( \phi; z ) &=& \frac{ F z }{ 2 } \phi^2 - \frac g4 \phi^4
\,.
\end{eqnarray}
The value $\phi = 0$ is unstable for $z \le 0$, as can be seen from the potentials sketched for a few values of $z$ in Fig.\:\ref{fig:potential}. 
The condensate amplitude is ``rolling up the hill'' and comes only for $z \to \infty$ to a standstill.

As is it familiar for a mechanical system, 
we have the energy conservation law
\begin{equation}
\frac{ {\rm d} }{ {\rm d}z }
H( \phi, p; z) =
\frac{ \partial }{ \partial z }
H( \phi, p; z) = 
\frac{F}{2} \phi^2
\,.
\label{eq:energy-conservation}
\end{equation}
At this point, it is essential that the potential is strictly linear.
By integrating both sides of Eq.\,(\ref{eq:energy-conservation}), 
we get
\begin{equation}
H( \phi(z), p(z); z) \bigg|_{z_1}^{z_2} =
\frac{F}{2}
\int\limits_{z_1}^{z_2}\!{\rm d}z\,
\phi^2( z )
\,.
\label{eq:conservation-law}
\end{equation}
Evaluate this for $z_1 \to -\infty$ and $z_2 = 0$:
\begin{equation}
H( \phi(0), p(0); 0) =
\frac{F}{2}
\int\limits_{-\infty}^{0}\!{\rm d}z\,
\phi^2( z )
= \frac{ F }{ 2 } N( z< 0 )
\,.
\label{eq:find-left-number}
\end{equation}
Indeed, the wave function vanishes in the tunnelling region
so that $H$ at the lower boundary does not contribute.
On the other half of the $z$-axis, 
subtract $(F z)^2 / 4g $ on both sides of Eq.\,(\ref{eq:conservation-law})
and re-write it in the form
\begin{eqnarray}
&& \Big[ H( \phi(z), p(z); z) - \frac{ (F z)^2 }{ 4g } \Big]_{0}^{z}
\nonumber\\
&& \qquad =
\frac{F}{2}
\int\limits_{0}^{z}\!{\rm d}z'
\Big( \phi^2( z' ) - \frac{ F z' }{ g } \Big)
\,.
\label{eq:number-positive-x}
\end{eqnarray}
When taking the limit $z \to \infty$, we enforce the Thomas-Fermi asymptotics
$\phi(z) \approx \sqrt{F z/g}$.
This gives a momentum $p \approx \frac12 \sqrt{F / g z} \to 0$ and 
a potential $V( \phi; z ) \approx (F z)^2 / 4g$. 
Since the post-Thomas-Fermi corrections vanish at infinity
\cite{Margetis00},
only the lower limit $z = 0$ contributes
on the left-hand side of Eq.\,(\ref{eq:number-positive-x}).
The right-hand side becomes the density difference defined in 
Eq.\,(\ref{eq:total-density-conserved}) so that we have
\begin{equation}
- H( \phi(0), p(0); 0) = \frac{ F }{ 2 } \delta N( z > 0 )
\,.
\label{eq:find-right-number}
\end{equation}
The desired result~(\ref{eq:total-density-conserved}) follows by adding
Eqs.\,(\ref{eq:find-left-number}, \ref{eq:find-right-number}).

\section{Fluctuation-induced particle density}
\label{a:n-prime-3rd-order}

In this Appendix, we discuss in more detail the fluctuation correction to the total density of the Bose gas. 
We follow the Bogoliubov theory for quasi-condensates of \citet{Mora03}.
This approach, technically speaking, has to push certain expansions in fluctuations around the condensate $\phi$ to the third order to get the next-to-leading order contribution to the density. 
This is why the expression provided by Eqs.\,(53, 65) of Ref.\:\onlinecite{Mora03} looks rather involved
\begin{equation}
n^{(2+3)} = 
\langle \hat{B}^\dag \hat{B} \rangle 
+ \langle \hat{P}^2 \rangle \phi_a^2 
- \phi \left( \phi_a + 2 \eta_c \right)
- 2 \phi \,\eta_B
\,.
\label{eq:65-Mora03-2}
\end{equation}
Here, the Bogoliubov operator $\hat{B}$ is defined by the mode expansion~(\ref{eq:uv-mode-expansion}).
Recalling Eq.\,(\ref{eq:fluctuation-operator}), the operator $\hat{P} \phi_a$ describes fluctuations in the condensate density.
The quantity $\eta_c$ is found as the solution to the following inhomogeneous equation ($H_3$ defined in Eq.\,(\ref{eq:H3-chi}))
\begin{equation}
H_3 \, \eta_c = \langle \hat{P}^2 \rangle 
(\mu'_0)^2 \Big(3 g \phi \,\frac{ \partial \phi }{ \partial \mu }
	- 1 \Big) \frac{ \partial \phi }{ \partial \mu }
	- \frac{ \mu'_0 }{2} g\phi^2 \frac{ \partial \phi }{ \partial \mu }
\,,
\label{eq:H3-eta-c-equal-Sc}
\end{equation}
where the derivative $\mu'_0$ of the chemical potential was defined in Eq.\,(\ref{eq:def-mu-prime-0}) and is sub-leading in the thermodynamic limit.
Finally, $\eta_B$ solves 
\begin{equation}
H_3 \, \eta_B = g \phi \,\langle 2 \hat{B}^\dag \hat{B}
	+ {\tfrac{1}{2}} \hat{B}^2 
	+ {\tfrac{1}{2}} \hat{B}^{\dag2} \rangle 
\,.
\label{eq:Mora03-for-etaB}
\end{equation}
This last contribution has been discussed in Secs.\:\ref{s:HFB-vs-Mora03}, \ref{s:IR-regular-subtraction} and describes the back-action on the condensate due to thermally excited quasiparticles.

To interpret the additional terms in Eq.\,(\ref{eq:65-Mora03-2}), Mora and Castin invoke particle fluctuations in the grand-canonical ensemble.
To see this, we need the second $\mu$-derivative of the Gross-Pitaevskii equation: 
after re-arranging terms, we find from Eq.\,(\ref{eq:H3-chi}) 
\begin{equation}
H_3 \, \frac{\partial^2 \phi }{ \partial \mu^2 } 
= 
- 6 g \phi \,\Big( \frac{ \partial \phi}{ \partial \mu } \Big)^2
+ 2 \frac{ \partial \phi}{ \partial \mu } 
\,.
\label{eq:H3-and-chi-2nd}
\end{equation}
$\eta_c$ in Eq.\,(\ref{eq:H3-eta-c-equal-Sc}) hence contains a piece proportional to $\partial^2 \phi / \partial \mu^2$:
\begin{equation}
\eta_c = - \tfrac{ 1 }{ 2 } \langle \hat{P}^2 \rangle (\mu'_0)^2 \, 
\frac{\partial^2 \phi }{ \partial \mu^2 }
- \frac{ \eta_a }{ 2 }
\,.
\label{eq:result-eta-c}
\end{equation}
The inhomogeneous equation for $\eta_a$ contains the remaining
term of Eq.\,(\ref{eq:H3-eta-c-equal-Sc}):
\begin{equation}
H_c \eta_a = \mu'_0 \, g \phi^2 \frac{ \partial \phi }{ \partial \mu }
\,.
\label{eq:H3-eta-a}
\end{equation}
Pulling these results together, we get for the second and third terms in 
Eq.\,(\ref{eq:65-Mora03-2}):
\begin{eqnarray}
n_c^{(2)}(z; T) &=& \langle \hat{P}^2 \rangle \phi_a^2 
 - \phi \left( \phi_a + 2 \eta_c \right)
\nonumber\\
& =& \mu'_0 T \Big( \dfrac{ \partial \phi}{ \partial \mu } \Big)^2
+ \mu'_0 T \, \phi \, \frac{\partial^2 \phi }{ \partial \mu^2 }
- \phi \left( \phi_a - \eta_a \right)
\nonumber\\
& =&
\frac{ \mu'_0 T }{ 2 }
\frac{\partial^2 \phi^2 }{ \partial \mu^2 }
- \phi \left( \phi_a - \eta_a \right)
\,.
\label{eq:result-n2-c}
\end{eqnarray}
As discussed in Ref.\:\onlinecite{Mora03}, the first term with the second derivative of the condensate density arises from the thermal fluctuations of condensate particles.
The second term is a quantum contribution related to the fact that the particle interaction energy is proportional to $\frac12 N(N-1)$ rather than $N^2/2$, see Appendix~A.3 in the supplementary material of Ref.\,\onlinecite{Henkel25b}. 
It features a back-action correction $\eta_a$ to the adjoint mode $\phi_a$ defined in Eq.\,(\ref{eq:def-phi-a}).
Solving Eq.\,(\ref{eq:H3-eta-a}) in the Thomas-Fermi approximation, we see that this correction is significant
\begin{equation}
	\eta_a 
\approx \frac{\mu'_0 }{ 2 } \frac{ \partial \phi }{ \partial \mu }
= \frac{ \phi_a }{ 2 }
\,.
\label{eq:solution-eta-a}
\end{equation}

\begin{figure}[bthp]
\centerline{%
\includegraphics*[height=0.47\columnwidth]{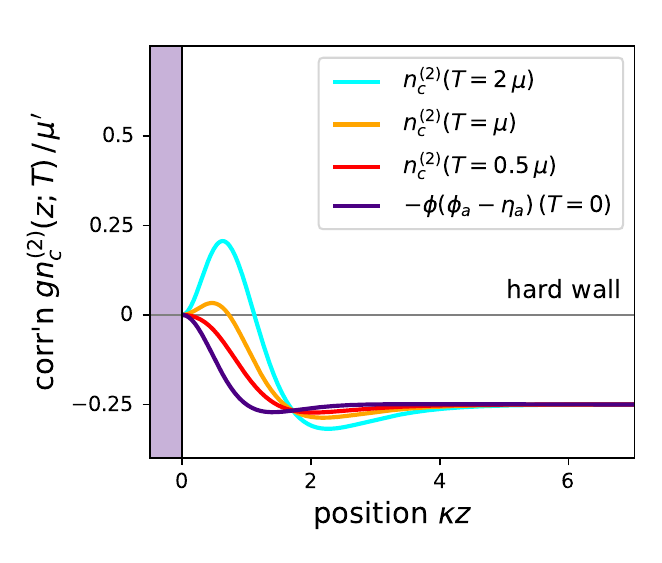}
\hspace*{-04mm}
\includegraphics*[height=0.47\columnwidth]{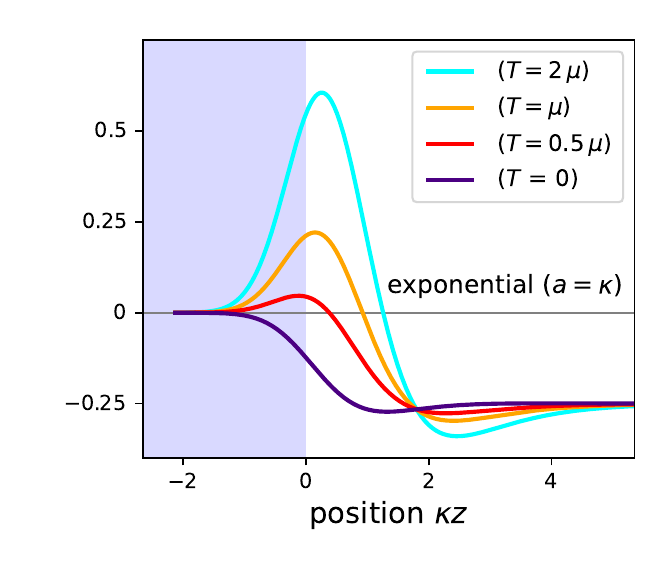}
}
\caption[]{Illustration of temperature-dependent corrections $n^{(2)}_c(z; T)$ to the particle density [Eq.\,(\ref{eq:result-n2-c})]. 
These terms arise from fluctuations in the particle number in the condensate and are negligible in the thermodynamic limit. 
They scale with the parameter $\mu'_0 \approx g/L$ when the region of interest is limited to $z \leq L$.
Left: hard wall; right: exponential potential with decay length $1/a = 1/\kappa$.
The curves increase in amplitude when $1/a$ gets larger (softer potential).
}
\label{fig:n-prime-GC}
\end{figure}

The two panels of Fig.\:\ref{fig:n-prime-GC} illustrate the temperature dependence of Eq.\,(\ref{eq:result-n2-c})
by comparing the hard wall and the exponential potential.
All four potentials we consider lead to curves with a similar shape.
Somewhat unusual is the quantitative difference when comparing  hard and soft wall potentials:
the overall amplitude is smaller for the hard-wall potential, although one may have expected the spatial variation of $\phi$ to be steeper.
The exponential potential ``wins'' because its density is not pinned at a fixed position: a variation of the chemical potential $\mu$ actually allows for a shift in the turning point, providing a larger change in the condensate mode.

\newcommand{\mybibpath}{/Users/carstenh/Work/BoxUP/documents/Biblio/}
\bibliography{\mybibpath journals,%
../../phase_shift}

\end{document}